\documentclass[a4paper]{article}

\pdfoutput=1

\usepackage{amssymb}
\usepackage{amsmath}
\usepackage{amssymb}
\usepackage{amstext}
\usepackage{mathtools}
\usepackage{graphicx}
\usepackage{fullpage}
\usepackage{array}
\usepackage{tabularx}
\usepackage{multirow}
\usepackage{longtable}
\usepackage{xcolor}

\usepackage{amsthm}

\usepackage{float}
\usepackage[skins,theorems]{tcolorbox}
\usepackage{authblk}

\usepackage[figuresright]{rotating}

\newcommand{\bs}[1]{\boldsymbol{#1}}
\def\bsb{\bs{b}}
\def\bsf{\bs{f}}
\def\bsu{\bs{u}}
\def\bsv{\bs{v}}

\providecommand{\keywords}[1]
{
  \small	
  \textbf{\textit{Keywords---}} #1
}

\begin{document}

\title{Valuation of the Convertible Bonds under Penalty TF model using Finite Element Method}

\author[1]{Rakhymzhan Kazbek}
\author[2,*]{Yogi Erlangga}
\author[1]{Yerlan Amanbek}
\author[1]{Dongming Wei}

\affil[1]{Nazarbayev University, Department of Mathematics, School of Sciences and Humanities, 53 Kabanbay Batyr Ave, Nur-Sultan 010000, Kazakhstan}
\affil[2]{Zayed University, Department of Mathematics, College of Natural and Health Sciences, Abu Dhabi Campus, P.O. Box 144534, United Arab Emirates}
\affil[ ]{\textit{rakhymzhan.kazbek@nu.edu.kz, yogi.erlangga@zu.ac.ae, yerlan.amanbek@nu.edu.kz, dongming.wei@nu.edu.kz}}
\affil[*]{Corresponding author}

\date{ }
\maketitle

\begin{abstract}
In this paper, the TF system of two-coupled Black-Scholes equations for pricing the convertible bonds is solved numerically by using the  P1 and P2 finite elements with the inequality constraints approximated by the penalty method. The corresponding finite element ODE system is numerically solved by using a modified Crank-Nicolson scheme, in which the non-linear system is solved at each time step by the Newton-Raphson method for non-smooth functions. Moreover, the corresponding Greeks are also calculated by taking advantage of the P1-P2 finite element approximation functions. Numerical solutions by the  finite element method compare favorably with the solutions by the finite difference method in literature.  
\end{abstract}


\section{Introduction}
\label{sec:intro}

Convertible bond (CB) is a type of financial derivatives used in financial risk management   and in trading by investors and the issuers ~\cite{hull2003options,wiley,handbook}. It has become a popular choice of corporations as a stable investment vehicle. 
Pricing financial derivatives for CBs is however a complicated problem. Tree methods (e.g., in~\cite{Milanov2012}), Monte-Carlo simulation (e.g., ~\cite{Ammann2008}), and partial differential equations are among the widely used techniques in the literature. 

One of the pioneering work using PDE modelling for CB pricing is due to Ignersoll~\cite{Jonatha-1977}, who developed a method for the determination of the optimal conversion and call policies for convertible securities using Black-Scholes~\cite{Black1973ThePO} methodology. Brennan and Schwartz~\cite{Brennan} extended Ignersoll's model to incorporate callability and dividend payments. They later included stochastic interest rate, resulting in a two-factor model for CB pricing with callability and conversion strategies~\cite{Brennan2}. In the late 1990s, Tsiveriotis and Fernandes~\cite{vcb} proposed an innovative two-factor model for pricing CBs under callability, puttability and conversion provisions with {\it hypothetical} cash-only convertible bond (COCB) part, known as the TF model. Another model for pricing CBs with credit risk and default strategy was developed by Ayache et al.~\cite{Ayache2003} based on a  system of triple partial differential equations, called the AFV model. 

Because of their complexity, the TF and AFV models have to be solved numerically. The review paper of Al Saedi and Tularam~\cite{AlSaedi2018} provides a recent account on methodologies for option pricing under the Black-Scholes equation, such as  the R3C scheme~\cite{Ankudinova2008}, cubic spline wavelets and multiwavelet bases methods~\cite{ern2016}, finite difference methods (FDM)~\cite{Dremkova2011}, finite-volume methods (FVM)~\cite{Lin2020}, and finite-element methods (FEM) \cite{Forsyth1999, BaroneAdesi2003, deFrutos2005, Kovalov2008}. Due to their simplicity, FDM have become a popular numerical method in computational finance. In computational bond pricing, the authors of ~\cite{Brennan, Brennan2, vcb, Ayache2003} use FDM to solve their proposed model.

In this paper, we discuss and expanded the existing literature on numerical solutions of the TF model for CB pricing with credit risk and without dividends by implementing finite-element methods.
The TF model for convertible bond pricing is based on the system of partial differential equations (PDEs)
\begin{align}
&\displaystyle \frac{\partial U}{\partial t}+ \frac{\sigma^2}{2}S^2\frac{\partial^2 U}{\partial S^2} +r_gS\frac{\partial U}{\partial S}-r(U-V)-(r+r_c)V=0,\label{Ueqxn}\\
&\displaystyle \frac{\partial V}{\partial t}+\frac{\sigma^2}{2}S^2\frac{\partial^2 V}{\partial S^2} +r_gS\frac{\partial V}{\partial S}-(r+r_c)V=0, \label{Beqxn}
\end{align}
for the time $t\in (0,T)$ and the underlying stock price $S \in (0,\infty)$, with $U$ being the value of the CB, $V$  the value of hypothetical COCB, $r$ the risk-free rate, $r_g$ the growth rate, which can be counted as risk-free rate $r$ (see~\cite{hull2003options}), $r_c$  the credit spread reflecting payoff default risk,   $\sigma$ be volatility, and $T$ the maturity time. 

The terminal condition at the maturity time $T$ means that once the CB is expired,  no one can call or put it back. Holder of the CB will get as much as possible depending on the conversion ratio $k$  and stock price. There is however a minimum based on the face value $F$ and the coupon payment $K$, yielding the terminal condition:
\begin{eqnarray}
U(S,T) = 
\begin{cases}
F+K, &\text{if } F+K\geq kS,\\
kS, &\text{otherwise}, 
\end{cases} \label{TcondU} 
\end{eqnarray}
and
\begin{eqnarray}
V(S,T) = 
\begin{cases}
F+K, &\text{if } F+K\geq kS,\\
0,&\text{otherwise}.
\end{cases} \label{TcondV} 
\end{eqnarray}
Throughout its lifetime, however, the CB can be converted to the underlying stock at the value $kS$, and issuer should pay the principal value $F+K$ to the holder, if the issuer has not converted until maturity time. These rights lead to three conditions that constrain the CB price:
\begin{enumerate}
\item Upside constraint due to conversion of bonds: for $t \in [0,T]$,
\begin{eqnarray}
&U(S,t)\geq kS, \label{eq:const_conv1} \\ 
&V(S,t) = 0 \text{ if } U(S,t)\leq kS; \label{eq:const_conv2}
\end{eqnarray}

\item  Upside constraint due to callability with the call price $B_{call}$: for $t \in [T_{call},T]$, with $T_{call}$ the earliest time the bond issuer is allowed to call back the bond, 
\begin{eqnarray}
&U(S,t)\leq \max(B_{call}(t),kS),\label{eq:const_up1} \\
&\quad\quad V(S,t) = 0 \text{ if } U(S,t)\geq B_{call}(t); \label{eq:const_up2}
\end{eqnarray}

\item Downside constraint due to putability with the put price $B_{put}$: for $t \in [T_{put},T]$, with $T_{put}$ the earliest time the investor is allowed to put the bond back,
\begin{eqnarray}
&U(S,t)\geq B_{put}(t), \label{eq:const_down1} \\
& V(S,t) = B_{put}(t) \text{ if } U(S,t)\leq B_{put}(t). \label{eq:const_down2}
\end{eqnarray}
\end{enumerate}
Following~\cite{Ayache2003}, the call and put price in the callability and puttability constraints include the effect of future's coupon payment and the underlying interest as follows. Let $\mathcal{T}_{\text{coupon}} = \{t_i\}$, the set of the coupon payment time, with $0 < t_{i-1} < t_i \le T$. Then
\begin{eqnarray}
   B_{put,call}(t) = B_{put,call}^{cl} + AccI(t),
\end{eqnarray}
where
\begin{eqnarray}
   AccI(t) = K_i \frac{t - t_{i-1}}{t_i - t_{i-1}},
\end{eqnarray}
the accrued interest at any time $t$ between the time of the last coupon payment $t_{i-1}$ and the time of the next (pending) coupon payment $t_i$. Note that the constraints~\eqref{eq:const_conv1} and~\eqref{eq:const_down1} can be combined to $U(S,t) \ge \max\{B_{put}, kS\}$. In this way, the constraints for $U$ can be rewritten as
\begin{align}
   U(S,t) &\le \max(B_{call}(t), kS), \\
   U(S,t) &\ge \max(B_{put}(t), kS),
\end{align}
with
\begin{align}
    B_{call}(t) &= \begin{cases}
             B_{call}^{cl} + AccI(t),&\text{if } t \in [T_{call},T], \\
             +\infty, &\text{otherwise},
          \end{cases} \\
    B_{put}(t) &= \begin{cases}
             B_{put}^{cl} + AccI(t),&\text{if } t \in [T_{put},T], \\
             0, &\text{otherwise}.
          \end{cases}
\end{align}

Two boundary conditions need to be supplemented to the PDEs~\eqref{Ueqxn} and~\eqref{Beqxn}. At $S = 0$, the PDEs are reduced to
\begin{align}
\displaystyle \frac{\partial U(0,t)}{\partial t} &=  rU(0,t) + r_c V(0,t), \label{bcond0}
\end{align}
and
\begin{align}
\displaystyle \frac{\partial V(0,t)}{\partial t} &= (r+r_c) V(0,t),
\label{bcond1}
\end{align}
with the terminal conditions $U(0,T) = V(0,T) = F+K$ (see Eqs. \eqref{TcondU} and \eqref{TcondV}). 
The other boundary condition is associated with the situation when the the stock price $S$ increases unboundedly, under which the CB is converted into stock. Therefore,
\begin{eqnarray}
\lim_{S \to \infty}
\begin{cases}
U(S,t)=kS, \\
V(S,t)=0.
\end{cases} \label{bcond1}
\end{eqnarray}

After spatial discretization using, for instance, FDM or FEM, the initial-boundary value problems (IBVP) with constraints given by~\eqref{Ueqxn}--\eqref{bcond1} can be solved by using a time-integration method, such as Crank-Nicolson, in combination with projected SSOR (PSSOR) method to tackle the nonlinearity. In our paper, however, we shall consider a formulation of the above-stated IBVP in a penalty PDE to explicitly include some constraints in the PDEs, and apply FEM and Crank-Nicolson method on the resulting penalty PDE.

The paper is organized as follows. In Section~\ref{sec:transform},  transformation of the TF model with penalty in a suitable form for numerical method is presented to implement our numerical schemes. Finite element formulation for the penalty TF model is established in Section~\ref{sec:FEM}. We detail the time-integration method for solving the resulting differential algebraic equation in Section~\ref{sec:timeintegral}. Section~\ref{sec:numresults} presents numerical results and discussions on the convergence of the FEM method. We finish up the paper by drawing some conclusion and remarks in Section~\ref{sec:conclusion}.

\section{The transformed TF model}
\label{sec:transform}

The standard procedure for solving the Black-Scholes-type PDEs requires transformation of the terminal-boundary value problem to an initial-boundary value problem, on which many numerical methods can be devised. Let $\displaystyle \tau = T - t $ and $\displaystyle x = \ln\left(S/S_{\text{int}}\right)$, 
where $S_{\text{int}}$ is the stock price at the initial time $t = 0$. With this change of variables, it can be shown that the PDEs \eqref{Ueqxn}--\eqref{Beqxn} are transformed to
\begin{align}
\displaystyle\frac{\partial U}{\partial \tau} &= \frac{\sigma^2}{2}\frac{\partial^2 U}{\partial x^2} +\left(r- \frac{\sigma^2}{2}\right)\frac{\partial U}{\partial x}-r(U - V) - (r + r_c)V, \label{eq:transU} \\
\displaystyle\frac{\partial V}{\partial \tau} &= \frac{\sigma^2}{2}\frac{\partial^2 V}{\partial x^2} + \left(r- \frac{\sigma^2}{2}\right)\frac{\partial V}{\partial x} - (r + r_c)V,
\label{eq:transV}
\end{align}
with $(\tau,x) \in (0,T)\times (-\infty,\infty)$. 
The terminal conditions at $t = T$ now become  initial conditions at $\tau = 0$, given by
\begin{eqnarray}
U(x,0) = 
\begin{cases}
F+K, &\text{if } F+K\geq kS_{\text{int}} e^x, \\
kS_{\text{int}} e^x , &\text{otherwise},
\end{cases} \label{TcondUtrans} 
\end{eqnarray}
and
\begin{eqnarray}
V(x,0) = 
\begin{cases}
F+K , &\text{if } F+K\geq kS_{\text{int}}e^x, \\
0 , &\text{otherwise}.
\end{cases}  \label{TcondVtrans} 
\end{eqnarray}
Furthermore, the constraints for $U$ are transformed into
\begin{align}
   U(x,\tau) &\ge \max\{B_{put} (\tau), kS_{\text{int}} e^x\} =: U_{put}^{\star}(\tau), \label{eq:tranformconstU1} \\
   U(x,\tau) &\le \max\{B_{call}(\tau), kS_{\text{int}} e^x\} =: U_{call}^{\star}(\tau), \label{eq:tranformconstU2}
\end{align}
where
\begin{align}
    B_{call}(\tau) &= \begin{cases}
             B_{call}^{cl} + AccI(\tau),&\text{if } t \in [0,\tau_{call}], \\
             +\infty, &\text{otherwise},
          \end{cases}
\end{align}
and
\begin{align}
    B_{put}(\tau) &= \begin{cases}
             B_{put}^{cl} + AccI(\tau),&\text{if } t \in [0, \tau_{put}], \\
             0, &\text{otherwise},
          \end{cases}
\end{align}
with $\tau_{put,call} = T - T_{put,call}$ and, for $\tau_i \in \mathcal{T}_{\tau, \text{coupon}} = \mathcal{T}_{\text{coupon}}$,
\begin{eqnarray} 
\displaystyle B_{put,call}(\tau) = B^{cl}_{put,call} + K_i\frac{\tau - \tau_{i-1}}{\tau_i - \tau_{i-1}}. \label{eqbcbptau}
\end{eqnarray}
The constraints for $V$ now read
\begin{enumerate}
\item for $\tau \in [0,T]$,
\begin{align}
    V(x,\tau) = 0, \text{ if } U(\tau)\leq kS_{\text{int}}e^x,
\end{align}
\item for $\tau \in [0,\tau_{call}]$,
\begin{align}
    V(x,\tau) = 0, \text{ if }  U(\tau) \geq B_{call}(\tau)
\end{align}
\item for $\tau \in [0,\tau_{put}]$,
\begin{align}
    V(x,\tau) = B_{put}, \text{ if }  U(\tau)\leq B_{put}(\tau). 
\end{align}
\end{enumerate}

For the boundary conditions, we note here that $x$ is not defined  at $S = 0$. As in the actual numerical computation we set $S$ as close as possible to $0$, we assume that~\eqref{bcond0} also holds at the proximity of $S = 0$, which corresponds to $x_{\min} \to - \infty$ in the $x$-space. This results in the boundary conditions at $x_{\min}$
\begin{align}
\frac{\partial U(x_{\min},\tau)}{\partial \tau} &=  -rU(x_{\min},\tau) - r_c V(x_{\min},\tau), \label{bcondx00} \\
\frac{\partial V(x_{\min},\tau)}{\partial t} &= -(r+r_c) V(x_{\min},\tau). \label{bcondx01}
\end{align}
Transformation of the boundary conditions at $S \to +\infty$ is straightforward: at $x_{\max} \to +\infty$,
\begin{align}
U(x_{\max},\tau) &= kS_{int}e^{x_{\max}}, \label{bcondx10}\\
\displaystyle V(x_{\max},\tau) &=0. \label{bcondx11}
\end{align}

As stated in Section~\ref{sec:intro}, our focus in this paper is on the the penalty TF model. We therefore need to reformulate the model into a PDE model with penalty terms associated with some constraints. In particular, as in practice we are mainly concerned with the CB price $U$, not $V$, we shall reformulate the CB PDE~\eqref{eq:transU} with the associated constraints~\eqref{eq:tranformconstU1} and ~\eqref{eq:tranformconstU2} into a penalty PDE. To this end, note that the linear complementarity problem (LCP)  for~\eqref{eq:transU} 
with constraints~\eqref{eq:tranformconstU1} and ~\eqref{eq:tranformconstU2} is given by~\cite{Ayache2003}
\begin{equation}
        \left(\begin{array}{l}
    \mathcal{L}U -r_cV = 0\\
    U \ge \max(B_p, \kappa S_{\text{int}} e^x) \\
    U \le \max(B_c, \kappa S_{\text{int}} e^x) 
\end{array}\right)
\vee 
\left(\begin{array}{l}
    \mathcal{L}U -r_cV \le 0\\
    U = \max(B_p, \kappa S_{\text{int}} e^x) \\
    U \le \max(B_c, \kappa S_{\text{int}} e^x) 
\end{array}\right)
\vee 
\left(\begin{array}{l}
    \mathcal{L}U -r_cV \ge 0\\
    U \ge \max(B_p, \kappa S_{\text{int}} e^x) \\
    U = \max(B_c, \kappa S_{\text{int}} e^x) 
\end{array}\right),
\label{eq:LCP1}
\end{equation} 
where $\mathcal{L} = -\displaystyle\frac{\partial} {\partial \tau} + \frac{\sigma^2}{2}\frac{\partial^2 }{\partial x^2} +\left(r- \frac{\sigma^2}{2}\right)\frac{\partial }{\partial x} - r$.

The penalty PDE for the bond valuation can be constructed from the LCP~\eqref{eq:LCP1}:
\begin{align}
    \frac{\partial U}{\partial \tau} = \frac{\sigma^2}{2}\frac{\partial^2 U}{\partial x^2} + \left(r- \frac{\sigma^2}{2} \right)\frac{\partial U}{\partial x}-rU  - r_cV + \rho\max(U - U_{call}^{\star},0)  + \rho\max(U_{put}^{\star} - U,0), \notag
\end{align}
where $\rho > 0$ is the penalty parameter, which typically is set very large. By rewriting $\max(U-U_{call}^{\star},0) = \alpha_{call} (U-U_{call}^{\star})$ and $\max(U_{put}^{\star}-U,0) = \alpha_{put} (U-U_{put}^{\star})$, with 
\begin{eqnarray}
    \alpha_{call} = \begin{cases}
                   1,& \text{if } U-U_{call}^{\star} \ge 0, \\
                   0,& \text{otherwise},
               \end{cases}
    \quad \text{and} \quad
    \alpha_{put} = \begin{cases}
                   1,& \text{if } U_{put}^{\star} -U\ge 0, \\
                   0,& \text{otherwise},
               \end{cases}
\end{eqnarray}
the CB PDE can be reformulated into
\begin{eqnarray}
    \frac{\partial U}{\partial \tau} = \frac{\sigma^2}{2}\frac{\partial^2 U}{\partial x^2}+(r- \frac{\sigma^2}{2})\frac{\partial U}{\partial x}-rU  - r_cV + \rho \alpha_{call} (U-U_{call}^{\star})  + \rho \alpha_{put} (U_{put}^{\star}-U). \label{eq:transpenU}
\end{eqnarray}

\section{Finite element method}
\label{sec:FEM}

In this section, we construct finite-element methods to approximately discretize the penalty CB PDE~\eqref{eq:transpenU} and the COCB PDE~\eqref{eq:transV} in the transformed TF model. To apply finite element spatial discretization to the PDEs, the domain $(0,\infty)$ is approximated by the bounded domain $\Omega=(x_{\min},x_{\max})$ in the following:

Suppose that the solution of the bond PDEs are functions of the following class:
$$
U(x,\tau),V(x,\tau) \in \mathcal{H}^1 = \left\{ f: \Omega \to \mathbb{R}  \mid f, \frac{\partial f}{\partial x} \in L_2(\Omega) \text{ and $f$ satisfies boundary conditions on } \partial \Omega \right\}.$$
Consider two test functions $w,z \in \mathcal{H}^1_0$, where $\mathcal{H}^1_0 = \left\{f:\Omega\to \mathbb{R} \mid f, \frac{\partial f}{\partial x} \in L_2(\Omega) \text{ and } f|_{\partial \Omega} = 0 \right\}$. The weak formulation of the penalty PDE~\eqref{eq:transpenU} and~\eqref{eq:transV} reads
\begin{equation}
\begin{aligned}
   \int\displaylimits_{\Omega}w\frac{\partial U}{\partial \tau}  &= \frac{\sigma^2}{2}\int\displaylimits_{\Omega}w\frac{\partial^2 U}{\partial x^2}  +\left(r- \frac{\sigma^2}{2}\right)\int\displaylimits_{\Omega}w\frac{\partial U}{\partial x} -r\int\displaylimits_{\Omega}wU  -r_c\int\displaylimits_{\Omega}wV  + \rho \int\displaylimits_{\Omega} \alpha_{call} w (U-U^{\star}_{call})   + \rho \int\displaylimits_{\Omega} \alpha_{put} w (U^{\star}_{put} -U), \notag \\
   \int\displaylimits_{\Omega}z\frac{\partial V}{\partial \tau}   &= \frac{\sigma^2}{2}\int\displaylimits_{\Omega}z\frac{\partial^2 V}{\partial x^2} + \left(r- \frac{\sigma^2}{2}\right)\int\displaylimits_{\Omega}z\frac{\partial V}{\partial x} - (r + r_c)\int\displaylimits_{\Omega}zV, \notag
\end{aligned}
\end{equation}
where the integration is carried out along the $x$-direction ($dx$ is not indicated to save space).
Integration by parts and applying the vanishing property of the function $w$ and $z$ at the boundary results in the weak formulation
\begin{equation}
\begin{aligned}
  \frac{\partial}{\partial \tau} \int\displaylimits_{\Omega}wU &= -\frac{\sigma^2}{2}\int\displaylimits_{\Omega}\frac{\partial w}{\partial x}\frac{\partial U}{\partial x}  - \left(r- \frac{\sigma^2}{2}\right)\int\displaylimits_{\Omega}\frac{\partial w}{\partial x}U-r\int\displaylimits_{\Omega}wU -r_c\int\displaylimits_{\Omega}wV + \rho \int\displaylimits_{\Omega} \alpha_{call} w (U - U^{\star}_{call})  + \rho \int\displaylimits_{\Omega} \alpha_{put} w (U^{\star}_{put} - U),  \notag \\
   \frac{\partial }{\partial \tau}\int\displaylimits_{\Omega}zV  &= -\frac{\sigma^2}{2}\int\displaylimits_{\Omega}\frac{\partial z}{\partial x}\frac{\partial V}{\partial x} - \left(r- \frac{\sigma^2}{2}\right)\int\displaylimits_{\Omega}\frac{\partial z}{\partial x}V - (r + r_c)\int\displaylimits_{\Omega}zV. \notag
\end{aligned}
\end{equation}

To build our finite-element approximation, we consider the finite dimensional subspace $S^h_0 \subset \mathcal{H}^1_0$, spanned by the basis $\{\psi_1,\psi_2,\dots,\psi_n\}$. The finite-element approximation to the solution $U$ is the function
$$
  U_h = \sum_{i=1}^n u_i \psi_i + \sum_{i \in \mathcal{I}_{\partial}} u_i \psi_i \simeq U,  \quad u_i \in \mathbb{R},
$$
where $\psi_{i \in \mathcal{I}_{\partial}}$ are additional functions needed to interpolate the given solutions at the boundaries. A similar form of approximation to $V$ is devised, namely
$$
  V_h = \sum_{i=1}^n v_i \psi_i + \sum_{i \in \mathcal{I}_{\partial}} v_i \psi_i \simeq V,  \quad v_i \in \mathbb{R}.
$$
The use of the above approximations results in the weak formulation in the finite-dimensional space:
\begin{equation}
\begin{aligned}
\frac{\partial}{\partial \tau} \left( \sum_{i=1}^n u_i \int\displaylimits_{\Omega} w \psi_i + \sum_{i\in \mathcal{I}_{\partial}} u_i \int\displaylimits_{\Omega} w \psi_i\right) &= -\frac{\sigma^2}{2} \left( \sum_{i=1}^n u_i  \int\displaylimits_{\Omega}\frac{\partial w}{\partial x}\frac{\partial \psi_i}{\partial x}  + \sum_{i\in\mathcal{I}_{\partial}} u_i  \int\displaylimits_{\Omega}\frac{\partial w}{\partial x}\frac{\partial \psi_i}{\partial x} \right) \\
&- \left(r- \frac{\sigma^2}{2}\right) \left( \sum_{i=1}^n u_i \int\displaylimits_{\Omega}\frac{\partial w}{\partial x}\psi_i   + \sum_{i\in\mathcal{I}_{\partial}}^n u_i \int\displaylimits_{\Omega}\frac{\partial w}{\partial x}\psi_i \right) \\
&-r \left( \sum_{i=1}^n u_i \int\displaylimits_{\Omega}w\psi_i + \sum_{i\in \mathcal{I}_{\partial}} u_i \int\displaylimits_{\Omega}w\psi_i \right)  - r_c \left( \sum_{i=1}^n v_i \int\displaylimits_{\Omega}w \psi_i + \sum_{i\in \mathcal{I}_{\partial}} v_i \int\displaylimits_{\Omega}w\psi_i \right) \\
&+ \mathcal{P}_{call} + \mathcal{P}_{put},
\end{aligned}
\end{equation}
where $\displaystyle \mathcal{P}_{call} = \rho  \int\displaylimits_{\Omega} \alpha_{call} w \left(U - U_{call}^{\star} \right) dx$ and $\displaystyle \mathcal{P}_{put} = \rho  \int\displaylimits_{\Omega} \alpha_{put} w \left(U_{put}^{\star} - U\right) dx$, 
and 
\begin{equation}
\begin{aligned}
\frac{\partial}{\partial \tau} \left( \sum_{i=1}^n v_i \int\displaylimits_{\Omega} z \psi_i + \sum_{i\in \mathcal{I}_{\partial}} v_i \int\displaylimits_{\Omega} z \psi_i\right) &= -\frac{\sigma^2}{2} \left( \sum_{i=1}^n v_i  \int\displaylimits_{\Omega}\frac{\partial z}{\partial x}\frac{\partial \psi_i}{\partial x}  + \sum_{i\in\mathcal{I}_{\partial}} v_i  \int\displaylimits_{\Omega}\frac{\partial z}{\partial x}\frac{\partial \psi_i}{\partial x} \right) \\
&- \left(r- \frac{\sigma^2}{2}\right) \left( \sum_{i=1}^n v_i \int\displaylimits_{\Omega}\frac{\partial z}{\partial x}\psi_i   + \sum_{i\in\mathcal{I}_{\partial}}^n v_i \int\displaylimits_{\Omega}\frac{\partial z}{\partial x}\psi_i \right) \\
&-(r+r_c) \left( \sum_{i=1}^n v_i \int\displaylimits_{\Omega}z\psi_i + \sum_{i\in \mathcal{I}_{\partial}} v_i \int\displaylimits_{\Omega}z\psi_i \right).
\end{aligned}
\end{equation}

In the Galerkin method~\cite{Kythe-2004}, the test function $w$ and $z$ are chosen to coincide with the basis function $\psi_i$. Imposing this condition for $w,z = \psi_j$, $j=1,\dots,n$ results in the system of equations
\begin{equation}
\begin{aligned}
\frac{\partial}{\partial \tau} \left( \sum_{i=1}^n u_i \int\displaylimits_{\Omega} \psi_j \psi_i + \sum_{i\in \mathcal{I}_{\partial}} u_i \int\displaylimits_{\Omega} \psi_j \psi_i\right) &= -\frac{\sigma^2}{2} \left( \sum_{i=1}^n u_i  \int\displaylimits_{\Omega}\frac{\partial \psi_j}{\partial x}\frac{\partial \psi_i}{\partial x}  + \sum_{i\in\mathcal{I}_{\partial}} u_i  \int\displaylimits_{\Omega}\frac{\partial \psi_j}{\partial x}\frac{\partial \psi_i}{\partial x} \right) \\
&- \left(r- \frac{\sigma^2}{2}\right) \left( \sum_{i=1}^n u_i \int\displaylimits_{\Omega}\frac{\partial \psi_j}{\partial x}\psi_i   + \sum_{i\in\mathcal{I}_{\partial}}^n u_i \int\displaylimits_{\Omega}\frac{\partial \psi_j}{\partial x}\psi_i \right) \\
&-r \left( \sum_{i=1}^n u_i \int\displaylimits_{\Omega}\psi_j\psi_i + \sum_{i\in \mathcal{I}_{\partial}} u_i \int\displaylimits_{\Omega} \psi_j \psi_i \right)  - r_c \left( \sum_{i=1}^n v_i \int\displaylimits_{\Omega}\psi_j \psi_i + \sum_{i\in \mathcal{I}_{\partial}} v_i \int\displaylimits_{\Omega}\psi_j \psi_i \right) \\
&+ \mathcal{P}_{call,j} + \mathcal{P}_{put,j},
\end{aligned}
\label{FEMsystem_of_u}
\end{equation}
where $\mathcal{P}_{call,j}$ is $\mathcal{P}_{call}$ with $w$ be replaced by $\psi_j$ and similarly for $\mathcal{P}_{put,j}$, and
\begin{equation}
\begin{aligned}
\frac{\partial}{\partial \tau} \left( \sum_{i=1}^n v_i \int\displaylimits_{\Omega} \psi_j \psi_i + \sum_{i\in \mathcal{I}_{\partial}} v_i \int\displaylimits_{\Omega} \psi_j \psi_i\right) &= -\frac{\sigma^2}{2} \left( \sum_{i=1}^n v_i  \int\displaylimits_{\Omega}\frac{\partial \psi_j}{\partial x}\frac{\partial \psi_i}{\partial x}  + \sum_{i\in\mathcal{I}_{\partial}} v_i  \int\displaylimits_{\Omega}\frac{\partial \psi_j}{\partial x}\frac{\partial \psi_i}{\partial x} \right) \\
&- \left(r- \frac{\sigma^2}{2}\right) \left( \sum_{i=1}^n v_i \int\displaylimits_{\Omega}\frac{\partial \psi_j}{\partial x}\psi_i   + \sum_{i\in\mathcal{I}_{\partial}}^n v_i \int\displaylimits_{\Omega}\frac{\partial \psi_j}{\partial x}\psi_i \right) \\
&-(r+r_c) \left( \sum_{i=1}^n v_i \int\displaylimits_{\Omega}\psi_j\psi_i + \sum_{i\in \mathcal{I}_{\partial}} v_i \int\displaylimits_{\Omega}\psi_j\psi_i \right).
\end{aligned}
\label{FEMsystem_of_v}
\end{equation}

In practice, systems of equations~\eqref{FEMsystem_of_u} and~\eqref{FEMsystem_of_v} are constructed via an assembly process using (local) element matrices, whose structures depend on the choice of the basis functions $\psi_i$. The common choice for the basis functions is a class of functions satisfying the nodal condition
\begin{eqnarray}
    \psi_i(x_j) = \begin{cases}
                         1,& i = j, \\
                         0,&\text{otherwise},
                    \end{cases} \label{eq:nodeq}
\end{eqnarray}
where $x_j$ is the nodal point. This choice leads to global systems of linear equations with sparse and banded coefficient matrices.

\subsection{Linear polynomial bases}

Consider partition of the spatial domain $\Omega$ into $n_E$ non-overlapping elements $\Omega_j = [x_{j-1},x_j]$, with $|\Omega_j| = x_j - x_{j-1} = h$, $x_j$, $j = 0,\dots, n_E$, the nodal points, $x_0 = x_{\min}$ and $x_{n_E} = x_{\max}$. In the basic element $\Omega_j = [x_{j-1},x_j]$, we define two linear interpolation basis functions
\begin{align}
    \psi_{j-1}(x) &= (x - x_j)/(x_{j-1} - x_j) = - (x-x_j)/h, \notag \\
    \psi_j (x)      &= (x-x_{j-1})/(x_j - x_{j-1}) = (x-x_{j-1})/h, \notag
\end{align}
resulting in the P1 finite element (P1-FEM). Evaluating the integrals in~\eqref{FEMsystem_of_u} and~\eqref{FEMsystem_of_v} using the above-stated basis functions over the element $\Omega_j$ results in the following local (element) matrices:
\begin{itemize}
\item For the $\displaystyle \int \psi_j \psi_i dx$ term, the element matrix reads
\begin{equation*}
    M_j =  \begin{bmatrix}
            \displaystyle  \int \displaylimits_{\Omega_{j}} \psi_{j-1} \psi_{j-1}dx & \displaystyle \int \displaylimits_{\Omega_{j}} \psi_{j-1} \psi_{j}dx \\
 \displaystyle \int \displaylimits_{\Omega_{j}} \psi_{j} \psi_{j-1}dx &\displaystyle  \int \displaylimits_{\Omega_{j}} \psi_{j} \psi_{j}dx
 \end{bmatrix} = \frac{h}{6} \begin{bmatrix}
    2 & 1 \\
    1 & 2
\end{bmatrix}.
\end{equation*}

\item For the $-\displaystyle \int \psi_{j,x} \psi_{i,x} dx$ term, the element matrix reads
\begin{equation*}
    K_j =  -\begin{bmatrix}
         \displaystyle  \int \displaylimits_{\Omega_{j}} \psi_{j-1,x} \psi_{j-1,x}dx & \displaystyle \int \displaylimits_{\Omega_{j}} \psi_{j-1,x} \psi_{j,x}dx \\
 \displaystyle \int \displaylimits_{\Omega_{j}} \psi_{j,x} \psi_{j-1,x}dx &\displaystyle  \int \displaylimits_{\Omega_{j}} \psi_{j,x} \psi_{j,x}dx
  \end{bmatrix} = 
  -\frac{1}{h} \begin{bmatrix}
   1 & -1 \\
 -1 & 1
  \end{bmatrix}.
\end{equation*}

\item For the $\displaystyle \int \psi_{j} \psi_{i,x} dx$ term, the element matrix reads
\begin{equation*}
    N_j =  \begin{bmatrix}
   \displaystyle  \int \displaylimits_{\Omega_{j}} \psi_{j-1} \psi_{j-1,x}dx & \displaystyle \int \displaylimits_{\Omega_{j}} \psi_{j-1} \psi_{j,x}dx \\
 \displaystyle \int \displaylimits_{\Omega_{j}} \psi_{j} \psi_{j-1,x}dx &\displaystyle  \int \displaylimits_{\Omega_{j}} \psi_{j} \psi_{j,x}dx
 \end{bmatrix} = 
 \frac{1}{2} \begin{bmatrix}
     -1 & -1 \\
      1 & 1
 \end{bmatrix}.
\end{equation*}
\end{itemize}

\subsection{Quadratic polynomial bases}

In this approach, we add a midpoint $x_{j-\frac{1}{2}} = (x_{j-1} + x_j)/2$ in the basic element $\Omega_j$, giving three nodal points: $x_{j-1}$, $x_{j-1/2}$, and $x_j$, and define three quadratic interpolation polynomials satisfying the nodal condition~\eqref{eq:nodeq}:
\begin{align}
    \psi_{j-1}(x) &= \frac{(x-x_{j-\frac{1}{2}})(x-x_j)}{(x_{j-1} - x_{j-\frac{1}{2}})(x_{j-1} - x_{j})} =  2(x-x_{j-\frac{1}{2}})(x-x_j)/h^2, \\
   \psi_{j-\frac{1}{2}}(x) &= \frac{(x-x_{j-1})(x-x_j)}{(x_{j-\frac{1}{2}} - x_{j-1})(x_{j-\frac{1}{2}} - x_{j})} = -4(x-x_{j-1})(x-x_j)/h^2, \\
    \psi_{j}(x) &= \frac{(x-x_{j-1})(x-x_{j-\frac{1}{2}})}{(x_j - x_{j-1})(x_j - x_{j-\frac{1}{2}})} =  2(x-x_{j-1})(x-x_{j-\frac{1}{2}})/h^2,
\end{align}
resulting in P2-FEM. The local element matrices are as follows:
\begin{itemize}
\item the $\displaystyle \int \psi_j \psi_i dx$ term:
\begin{equation*}
    M_j =  \begin{bmatrix}
     \displaystyle  \int \displaylimits_{\Omega_{j}} \psi_{j-1} \psi_{j-1}dx & \displaystyle \int \displaylimits_{\Omega_{j}} \psi_{j-1} \psi_{j-\frac{1}{2}}dx& \displaystyle \int \displaylimits_{\Omega_{j}} \psi_{j-1} \psi_{j}dx \\
 \displaystyle  \int \displaylimits_{\Omega_{j}} \psi_{j-\frac{1}{2}} \psi_{j-1}dx & \displaystyle \int \displaylimits_{\Omega_{j}} \psi_{j-\frac{1}{2}} \psi_{j-\frac{1}{2}}dx& \displaystyle \int \displaylimits_{\Omega_{j}} \psi_{j-\frac{1}{2}} \psi_{j}dx \\
 \displaystyle \int \displaylimits_{\Omega_{j}} \psi_{j} \psi_{j-1}dx  &\displaystyle  \int \displaylimits_{\Omega_{j}} \psi_{j} \psi_{j-\frac{1}{2}}dx &\displaystyle  \int \displaylimits_{\Omega_{j}} \psi_{j} \psi_{j}dx
\end{bmatrix} = 
\frac{h}{30} \begin{bmatrix}
   4 & 2 & -1 \\
   2 & 16 & 2 \\
  -1 & 2  & 4
\end{bmatrix}.
\end{equation*}
\item the $\displaystyle -\int \psi_{j,x} \psi_{i,x} dx$ term:
\begin{equation*}
    K_j = - \begin{bmatrix}
     \displaystyle  \int \displaylimits_{\Omega_{j}} \psi_{j-1,x} \psi_{j-1,x}dx & \displaystyle \int \displaylimits_{\Omega_{j}} \psi_{j-1,x} \psi_{j-\frac{1}{2},x}dx& \displaystyle \int \displaylimits_{\Omega_{j}} \psi_{j-1,x} \psi_{j,x}dx \\
 \displaystyle  \int \displaylimits_{\Omega_{j}} \psi_{j-\frac{1}{2},x} \psi_{j-1,x}dx & \displaystyle \int \displaylimits_{\Omega_{j}} \psi_{j-\frac{1}{2},x} \psi_{j-\frac{1}{2},x}dx& \displaystyle \int \displaylimits_{\Omega_{j}} \psi_{j-\frac{1}{2},x} \psi_{j,x}dx \\
 \displaystyle \int \displaylimits_{\Omega_{j}} \psi_{j,x} \psi_{j-1,x}dx  &\displaystyle  \int \displaylimits_{\Omega_{j}} \psi_{j,x} \psi_{j-\frac{1}{2},x}dx &\displaystyle  \int \displaylimits_{\Omega_{j}} \psi_{j,x} \psi_{j,x}dx
 \end{bmatrix} = 
 -\frac{1}{3h} \begin{bmatrix}
  7 & -8 & 1 \\
 -8 & 16 & -8 \\
 1 & -8  & 7
\end{bmatrix}.
\end{equation*}

\item the $\displaystyle \int \psi_{j} \psi_{i,x} dx$ term:
\begin{equation*}
   N_j = \begin{bmatrix}
        \displaystyle  \int \displaylimits_{\Omega_{j}} \psi_{j-1} \psi_{j-1,x}dx & \displaystyle \int \displaylimits_{\Omega_{j}} \psi_{j-1} \psi_{j-\frac{1}{2},x}dx& \displaystyle \int \displaylimits_{\Omega_{j}} \psi_{j-1} \psi_{j,x}dx \\
 \displaystyle  \int \displaylimits_{\Omega_{j}} \psi_{j-\frac{1}{2}} \psi_{j-1,x}dx & \displaystyle \int \displaylimits_{\Omega_{j}} \psi_{j-\frac{1}{2}} \psi_{j-\frac{1}{2},x}dx& \displaystyle \int \displaylimits_{\Omega_{j}} \psi_{j-\frac{1}{2}} \psi_{j,x}dx \\
 \displaystyle \int \displaylimits_{\Omega_{j}} \psi_{j} \psi_{j-1,x}dx  &\displaystyle  \int \displaylimits_{\Omega_{j}} \psi_{j} \psi_{j-\frac{1}{2},x}dx &\displaystyle  \int \displaylimits_{\Omega_{j}} \psi_{j} \psi_{j,x}dx
 \end{bmatrix} = \frac{1}{6} \begin{bmatrix}
  -3 & -4 & 1 \\
  4 & 0 & -4 \\
 -1 & 4  & 3
\end{bmatrix}.
\end{equation*}
\end{itemize}

\subsection{Treating the constraints by Penalty method}

We now turn to the two nonlinear penalty terms in~\eqref{eq:transpenU} and construct a finite-element approximation to them. We in particular apply {\it group} finite element~\cite{FLETCHER1983225} to deal with the nonlinearity. We shall discuss the construction for $\mathcal{P}_{call,j}$;  construction of finite element approximation for $\mathcal{P}_{put,j}$ is done in the same way.

We assume that the term $\zeta_{call} := \alpha_{call} (U-U^{\star}_{call})$ is approximated by
$$
  \zeta_{call} = \sum_{i=1}^n \zeta_i \psi_i + \sum_{\mathcal{I}_{\partial}} \zeta_i \psi_i,
$$
where $\zeta_{call,i} = \alpha_{call} (x_i) (U(x_i) - U^{\star}_{call}(x_i)) =: \alpha_{call,i}(u_i - u^{\star}_{call,i})$. Therefore, for $w = \psi_j$, $j = 1,\dots,n$, we have
\begin{align}
\displaystyle \mathcal{P}_{call,j} &= \rho  \int\displaylimits_{\Omega} \psi_j \left( \sum_{i=1}^n \zeta_i \psi_i + \sum_{\mathcal{I}_{\partial}} \zeta_i \psi_i \right) dx = \rho \left( \sum_{i=1}^n \zeta_i  \int\displaylimits_{\Omega} \psi_j  \psi_i dx + \sum_{\mathcal{I}_{\partial}} \zeta_i \int\displaylimits_{\Omega} \psi_j  \psi_i dx \right) \notag \\
&= \rho \left( \sum_{i=1}^n \alpha_{call,i}(u_i - u^{\star}_{call,i})  \int\displaylimits_{\Omega} \psi_j  \psi_i dx + \sum_{\mathcal{I}_{\partial}} \alpha_{call,i}(u_i - u^{\star}_{call,i})  \int\displaylimits_{\Omega} \psi_j  \psi_i dx \right).
\end{align}
Each integral in the above-equation is evaluated element-wise, resulting in the local element matrices $M_j$, $K_j$, and $N_j$, given in Sections 3.1 and 3.2.

By using the same argument,
\begin{align}
\displaystyle \mathcal{P}_{put,j} = \rho \left( \sum_{i=1}^n \alpha_{put,i}(u_i - u^{\star}_{put,i})  \int\displaylimits_{\Omega} \psi_j  \psi_i dx + \sum_{\mathcal{I}_{\partial}} \alpha_{put,i}(u_i - u^{\star}_{put,i})  \int\displaylimits_{\Omega} \psi_j  \psi_i dx \right).
\end{align}

\section{Time integration scheme}
\label{sec:timeintegral}

The global finite-element system obtained from assembling the local finite-element matrices can be represented by the differential algebraic equations (DAEs):
\begin{align}
    \frac{\partial}{\partial \tau}(M \bsu + \hat{\bsb}_{M,u})
    &= -\frac{\sigma^2}{2} K\bsu - \left(r - \frac{\sigma^2}{2} \right) N\bsu - rM\bsu - r_c 
    M\bsv - \boldsymbol{\beta}_1(\bsu,\bsv) \notag \\
    &+ \rho M P_{put} (\bsu_{put}^{\star} - \bsu) + \rho M P_{call} (\bsu - \bsu_{call}^{\star}) + \rho \bsb_{put} + \rho \bsb_{call}:= F_1(\bsu,\bsv), \label{eq:dae1} \\
    \frac{\partial}{\partial \tau}(M\bsv  + \hat{\bsb}_{M,v})
    &= -\frac{\sigma^2}{2}  K\bsv  - \left( r - \frac{\sigma^2}{2} \right) N \bsv - (r+r_c) M \bsv - \boldsymbol{\beta}_2 (\bsv) := F_2(\bsv), \label{eq:dae2}
\end{align}
where $P_{put} = \text{diag}(\alpha_{put,j})$, $P_{call} = \text{diag}(\alpha_{call,j})$, and
\begin{align}
    \displaystyle \boldsymbol{\beta_1}(\bsu,\bsv) &= \frac{\sigma^2}{2}  \bsb_{K,u} + \left(r - \frac{\sigma^2}{2}  \right)\boldsymbol{b}_{N,u}+r \bsb_{M,u}+ r_c\bsb_{M,v}, \\
    \displaystyle \boldsymbol{\beta}_2 (\bsv) &= \frac{\sigma^2}{2} \bsb_{K,v} + \left(r - \frac{\sigma^2}{2} \right)\bsb_{N,v} + (r+r_c)\bsb_{M,v},
\end{align}
are the boundary condition vectors.

Time integration of the DAEs~\eqref{eq:dae1} and~\eqref{eq:dae2} is carried out by applying the $\theta$-scheme on both equations, which results in the systems, with $\theta \in [0,1]$, $\Delta \tau = T/n_{\tau}$, and $n_{\tau}$ the number of time steps,
\begin{align}
  M\bsu^{m+1} + \hat{\bsb}_{M,u}^{m+1} - M\bsu^{m} - \hat{\bsb}_{M,u}^{m} = \theta \Delta\tau F_1(\bsu^{m+1},\bsv^{m+1}) + (1-\theta)\Delta\tau F_1(\bsu^{m},\bsv^{m}), \notag  \\
  M \bsv^{m+1} + \hat{\bsb}_{M,v}^{m+1} - M \bsv^{m} -\hat{\bsb}_{M,v}^{m} = \theta\Delta\tau F_2(\bsv^{m+1}) + (1-\theta)\Delta\tau F_2(\bsv^{m}), \notag
\end{align}
or
\begin{align}
    A_{11} \bsu^{m+1} &+ A_{12} \bsv^{m+1} - \rho \theta \Delta \tau M\left(P_{put}^{m+1} (\bsu_{put}^{\star,m+1} - \bsu^{m+1}) + P_{call}^{m+1}(\bsu^{m+1}-\bsu^{\star,m+1}_{call})\right) \notag \\
    &= \widetilde{A}_{11}\bsu^m + \widetilde{A}_{12} \bsv^m + \rho (1-\theta) \Delta \tau M\left(P_{put}^{m} (\bsu_{put}^{\star,m} - \bsu^{m}) + P_{call}^{m}(\bsu^{m}-\bsu^{\star,m}_{call})\right) \notag \\
    &+ \theta \Delta \tau \boldsymbol{\beta}_1^{m+1} +  (1-\theta) \Delta \tau \boldsymbol{\beta}_1^{m} + \hat{\bsb}^m_{M,u} - \hat{\bsb}^{m+1}_{M,u} + \theta \rho \Delta \tau (\bsb_{put}^{m+1} + \bsb_{call}^{m+1}) + (1-\theta) \rho \Delta \tau (\bsb_{put}^{m} + \bsb_{call}^{m}), \label{CNsystem1} \\
    A_{22} \bsv^{m+1} &= \widetilde{A}_{22}\bsv^m + \theta \Delta \tau \boldsymbol{\beta}_2^{m+1} +  (1-\theta) \Delta \tau \boldsymbol{\beta}_2^{m} + \hat{\bsb}^m_{M,v} - \hat{\bsb}^{m+1}_{M,v}, \label{CNsystem2}
\end{align}
where
\begin{align}
    A_{11} &= M + \theta \Delta \tau \left( \frac{\sigma^2}{2} K + \left(r - \frac{\sigma^2}{2}  \right)N + r M  \right), \notag \\
    A_{12} &= \theta \Delta \tau r_c M, \notag \\
    A_{22} &= M + \theta \Delta \tau \left( \frac{\sigma^2}{2} K + \left(r - \frac{\sigma^2}{2}  \right)N + (r+r_c) M  \right) \notag \\
    \widetilde{A}_{11} &= M - (1-\theta) \Delta \tau \left( \frac{\sigma^2}{2} K + \left(r - \frac{\sigma^2}{2}  \right)N + r M  \right), \notag \\
    \widetilde{A}_{12} &= -(1-\theta) \Delta \tau r_c M, \notag \\
    \widetilde{A}_{22} &= M - (1-\theta) \Delta \tau \left( \frac{\sigma^2}{2} K + \left(r - \frac{\sigma^2}{2}  \right)N + (r+r_c) M  \right). \notag
\end{align}

Let the solutions $\bsu^m$ and $\bsv^m$ be known. The solutions at the next time level $m+1$ can in principle be computed by first solving~\eqref{CNsystem2} for $\bsv^{m+1}$. The solution $\bsu^{m+1}$ is then computed via~\eqref{CNsystem1} using the known $\bsu^{m}$, $\bsv^{m}$, and $\bsv^{m+1}$. This procedure however requires knowledge of the solutions at the boundaries at the time level $m+1$.

\subsection{Boundary solutions}

At $x_{\min}$, with $u_0(\tau) := U(x_{\min},\tau)$, $v_0(\tau) := V(x_{\min},\tau)$, etc, the boundary conditions with penalty in $U$ can be written as follows:
\begin{eqnarray}
\begin{cases}
\displaystyle \frac{\partial u_0(\tau)}{\partial \tau} =  -ru_0(\tau) - r_c v_0(\tau) + \rho \max\left(u_0(\tau) - u^{\star}_{call,0}(\tau),0\right) + \rho \max\left(u^{\star}_{put,0}(\tau) - u_0(\tau),0\right),\\[10pt]
\displaystyle \frac{\partial v_0(\tau)}{\partial t} = -(r+r_c) v_0(\tau).
\end{cases} \label{bcond0a}
\end{eqnarray}
Note that $\max\left(u_0(\tau) - u^{\star} _{call,0}(\tau),0\right) = p_{call,0}(\tau) (u_0(\tau) - u^{\star}_{call,0}(\tau))$, with
$$
  p_{call,0} = \begin{cases}
                   1,& \text{if } u_0(\tau) > u^{\star}_{call,0}(\tau), \\
                   0,& \text{otherwise},
               \end{cases}
$$
and $\max\left(u^{\star}_{put,0}(\tau) - u_0(\tau),0\right) = p_{put,0}(\tau) (u^{\star}_{put,0}(\tau) - u_0(\tau))$, with 
$$
  p_{put,0} = \begin{cases}
                   1,& \text{if } u_0(\tau) < u^{\star}_{put,0}(\tau), \\
                   0,& \text{otherwise}.
               \end{cases}
$$
Application of the $\theta$-scheme on \eqref{bcond0a} leads to the discrete equations:
\begin{align}
    u_0^{m+1} + \theta \Delta \tau &\left( r u_0^{m+1} + r_c v_0^{m+1} + \rho \left(p_{call,0}^{m+1}(u_0^{m+1} - u_{call,0}^{\star,m+1}) + p_{put}^{m+1}(u_{put,0}^{\star,m+1} - u_0^{m+1}) \right) \right) \notag \\
    &= u_0^m - (1-\theta) \Delta \tau \left(r u_0^{m} + r_c v_0^{m} + \rho \left(p_{call,0}^{m}(u_0^{m} - u_{call,0}^{\star,m}) + p_{put}^{m}(u_{put,0}^{\star,m} - u_0^{m}) \right) \right), \label{CNbcond1} \\
    \left[1 + \theta \Delta \tau (r+r_c)\right] v_0^{m+1} &= [1 - (1-\theta) \Delta \tau (r+r_c)] v_0^m. \label{CNbcond2}
\end{align}

Let the boundary solution $v_0^m$ be known. Then $v_0^{m+1}$ can be computed from~\eqref{CNbcond2}. With $u_0^m$, $v_0^m$, and $v_0^{m+1}$ now known,~\eqref{CNbcond1} becomes a nonlinear function of $u_0^{m+1}$, which can be solved approximately using Newton's method. First of all we assume that the penalty term in~\eqref{CNbcond1} is approximated in a fully implicit way at the new time level $m+1$. This results in the equation
\begin{eqnarray}
  0 = (1 + \theta \Delta \tau r)u_0^{m+1} + \Delta \tau  \rho \left(p_{call,0}^{m+1}(u_0^{m+1} - u_{call,0}^{\star,m+1}) + p_{put}^{m+1}(u_{put,0}^{\star,m+1} - u_0^{m+1}) \right) - \phi_0 =: f(u_0^{m+1}), \label{f_bc}
\end{eqnarray}
where
$$
  \phi_0 = u_0^m - (1-\theta) \Delta \tau \left(r u_0^{m} + r_c v_0^{m} \right) - \theta \Delta \tau r_c v_0^{m+1}.
$$
With
\begin{eqnarray}
   f'(u_0^{m+1}) = 1 + \theta \Delta \tau r + \Delta \tau \rho \left(p_{call,0}^{m+1} - p_{put,0}^{m+1}\right), \label{fp_bc}
\end{eqnarray}
Newton's method for finding $u_0^{m+1}$ satisfying~\eqref{f_bc} can be written as follows: with an initial guess $u_0^{m+1,0}$, compute $u_0^{m+1,k} = u_0^{m+1,k-1} - f(u_0^{m+1,k-1})/f'(u_0^{m+1,k-1})$, for $k = 1,2,\dots$.

In this paper, $u_0^{m+1,0}$ is chosen to be solution of the unconstrained boundary problem given by~\eqref{bcond0}. Since we expect that $u_0^{m+1,0}$ computed in this way is a better approximation than, e.g., $u_0^m$, we can use this value as well to evaluate the conditions to constraint $v$.

The complete algorithm for computing $u_0^{m+1}$ and $v_0^{m+1}$ is as follows:\\
\newline
\noindent \underline{Algorithm 1: Computing the boundary solutions} 
\begin{enumerate}
    \item input $u_0^m$, $v_0^m$;
    \item compute $B_p(\tau^{m+1})$ and $B_c(\tau^{m+1})$;
    \item compute $v_0^{m+1}$ from~\eqref{CNbcond2};
    \item compute $u_0^{m+1}$ from~\eqref{CNbcond1} without penalty terms;
    \item apply constraints on $v_0^{m+1}$ using $u_0^{m+1}$;
    \item set $u_0^{m+1,0} \leftarrow u_0^{m+1}$;
    \item for $k = 1,2,\dots$ until convergence
    \begin{enumerate}
        \item[7.1.] compute $f(u_0^{m+1,k-1})$ using~\eqref{f_bc};
        \item[7.2.] compute $f'(u_0^{m+1,k-1})$ using~\eqref{fp_bc};
        \item[7.3.] $u_0^{m+1,k} \leftarrow  u_0^{m+1,k-1} - f(u_0^{m+1,k-1})/f'(u_0^{m+1,k-1})$;
    \end{enumerate}
    \item apply constraints on $v_0^{m+1}$ using $u_0^{m+1,k}$;
    \item if $\tau^{m+1} \in \mathcal{T}_{\text{coupon}}$
    \begin{enumerate}
        \item[9.1] $u_0^{m+1} \leftarrow u_0^{m+1} + K$;
        \item[9.2] $v_0^{m+1} \leftarrow v_0^{m+1} + K$;
    \end{enumerate}
\end{enumerate}

At $x_{\max}$ we need to compute $u_{n+1}^{m+1} := U(x_{\max},\tau^{m+1})$ and $v_{n+1}^{m+1} := V(x_{\max},\tau^{m+1})$ via~\eqref{bcond1}, apply the constraints, and pay the coupon if $\tau^{m+1} \in \mathcal{T}_{\text{coupon}}$.

We remark here that solution of Newton's method exists if the derivative $f'(u_0^{m}) \neq 0$. This condition however cannot be theoretically guaranteed. In this regard, for the tuple $(p_{call,0}^{m}, p_{put,0}^{m}) \in \{0,1\}^2$ and nonnegative $\theta$, $\Delta \tau$, $\rho$, and $r$,
\begin{enumerate}
\item if $p_{call,0} = p_{put,0}$, then $f'(u_0^m) = 1 + \theta \Delta \tau r > 0$;
\item if $p_{call,0} \neq p_{put,0}$, we have two situations:
\begin{enumerate}
    \item if $p_{call,0} = 1$ and $p_{put,0} = 0$, then $f'(u_0^m) = 1 + \Delta \tau (\theta r + \rho) > 0$;
    \item if $p_{call,0} = 0$ and $p_{put,0} = 1$, then $f'(u_0^m) = 1 + \Delta \tau (\theta r - \rho) = 0$ if $\Delta \tau = 1/(\rho - \theta r)$. 
\end{enumerate}
\end{enumerate}
Since the penalty parameter $\rho$ is taken very large (e.g., $\rho = 10^{12}$ in this paper), vanishing $f'$ is very unlikely to happen under a reasonable choice of the time step $\Delta \tau$. For instance, using the simulation parameters in Table~\ref{tab:parameter} in Section~\ref{sec:numresults}, the vanishing $f'$ situation outlined in Point 2(b) above may occur if $\Delta \tau$ is chosen to be in the order of $\rho^{-1} = 10^{-12}$, which makes the time-integration process  extremely impractical.

\subsection{Interior solutions}

With solutions at the boundaries available at $\tau^m$ and $\tau^{m+1}$, all related boundary vectors in~\eqref{CNsystem1} and~\eqref{CNsystem2} are known. We are thus now in the position to compute the solutions $\bsu^{m+1}$ and $\bsv^{m+1}$. $\bsv^{m+1}$ is readily computed from~\eqref{CNsystem2}. With $\bsv^{m+1}$ known,~\eqref{CNsystem1} reduces to a nonlinear function of $\bsu^{m+1}$. As we do for the boundary equations, we assume that the penalty terms in~\eqref{CNsystem2} are approximated in a fully implicit way, resulting in the equation
\begin{eqnarray}
   0 &= A_{11} \bsu^{m+1} - \rho \Delta \tau M\left(P_{put}^{m+1} (\bsu_{put}^{\star,m+1} - \bsu^{m+1}) + P_{call}^{m+1}(\bsu^{m+1}-\bsu^{\star,m+1}_{call})\right) 
    - \boldsymbol{\phi} := \boldsymbol{f}(\bsu^{m+1}), \label{eqnonlinU}
\end{eqnarray}
where
\begin{eqnarray}
   \boldsymbol{\phi} = \widetilde{A}_{11}\bsu^m + \widetilde{A}_{12} \bsv^m  - A_{12} \bsv^{m+1} + \theta \Delta \tau \boldsymbol{\beta}_1^{m+1} +  (1-\theta) \Delta \tau \boldsymbol{\beta}_1^{m} + \hat{\bsb}^m_{M,u} - \hat{\bsb}^{m+1}_{M,u} + \theta \Delta \tau (\bsb_{put}^{m+1} + \bsb_{call}^{m+1}).
\end{eqnarray}

The nonlinear equation~\eqref{eqnonlinU} is solved iteratively using Newton's method. Starting from an initial guess of the solution $\bsu^{m+1,0}$, the solution $\bsu^{m+1}$ is approximated using the iterands
$$
   \bsu^{m+1,k} \leftarrow \bsu^{m+1,k-1} - \left( \nabla \bsf (\bsu^{m+1,k-1})\right)^{-1} \bsf(\bsu^{m+1,k-1}), \quad k = 1, 2, \dots
$$
where $\nabla \bsf(\bsu^{m+1,k-1})$ is the Jacobian of $\bsf$, given by
\begin{eqnarray}
  \nabla \bsf(\bsu^{m+1,k-1}) = A_{11} + \rho \Delta \tau M \left(P_{call}^{m+1,k-1} - P_{put}^{m+1,k-1} \right).  \label{eqjacf}
\end{eqnarray}
The initial guess $\bsu^{m+1,0}$ is chosen such that it solves unconstrained CB PDE, which is equivalent to solving~\eqref{CNsystem1} without penalty terms. We shall also use this unconstrained solution $\bsu^{m+1,0}$ to constrain the initially computed $\bsv^{m+1}$ prior to the start of Newton's iterations. The procedure for computing the solutions in the interior after one $\theta$-scheme step is summarized in the following algorithm: \\
\newline
\noindent \underline{Algorithm 2: Computing the interior solutions} 
\begin{enumerate}
    \item input $\bsu^m$, $\bsv^m$;
    \item compute $B_p(\tau^{m+1})$ and $B_c(\tau^{m+1})$;
    \item compute $\bsv^{m+1}$ from~\eqref{CNsystem2};
    \item compute $\bsu^{m+1}$ from~\eqref{CNsystem1} without penalty terms;
    \item apply constraints on $\bsv^{m+1}$ using $\bsu^{m+1}$;
    \item set $\bsu^{m+1,0} \leftarrow \bsu^{m+1}$;
    \item for $k = 1,2,\dots$ until convergence
    \begin{enumerate}
        \item[7.1.] compute $\bsf(\bsu^{m+1,k-1})$ using~\eqref{eqnonlinU};
        \item[7.2.] compute $\nabla \bsf(\bsu^{m+1,k-1})$ using~\eqref{eqjacf};
        \item[7.3.] $\bsu^{m+1,k} \leftarrow \bsu^{m+1,k-1} - \left( \nabla \bsf(\bsu^{m+1,k-1}) \right)^{-1} \bsf(\bsu^{m+1,k-1})$;
    \end{enumerate}
    \item apply constraints on $\bsv^{m+1}$ using $\bsu^{m+1,k}$;
    \item if $\tau^{m+1} \in \mathcal{T}_{\text{coupon}}$
    \begin{enumerate}
        \item[9.1] $\bsu^{m+1} \leftarrow \bsu^{m+1} + K$;
        \item[9.2] $\bsv^{m+1} \leftarrow \bsv^{m+1} + K$;
    \end{enumerate}
\end{enumerate}

The existence of a solution of Newton's method requires nonsingularity of the Jacobian $\nabla f$. Like in the case with Newton's method for computing boundary solutions, there may exist values of parameters $r$, $h$, $\sigma$, $\theta$, $\Delta \tau$, and $\rho$ such that the Jacobian is singular. Quantification of such conditions requires analysis, which is beyond the scope of this paper. Numerical tests using various realistic values of parameters exhibit no convergence issues with the methods, suggesting nonsingularity of $\nabla f$.

\subsection{Summary of the time integration method}

By including Algorithm 1 and 2 in the $\theta$-scheme, the complete time stepping procedure to compute the solutions $\bsu(x,\tau)$ and $\bsv(x,\tau)$ is shown in Algorithm 3.\\
\newline
\noindent \underline{Algorithm 3: $\theta$-scheme for time integration with constraints}
\begin{enumerate}
    \item input the initial solution at $\tau^0 = 0$: $\bsu^0$, $\bsv^0$, $u_0^0$, $u_{n+1}^0$, $v_0^0$, and $v_{n+1}^0$;
    \item for $m = 0,1, \dots, n_{\tau} - 1$
    \begin{enumerate}
        \item[2.1.] Compute $B_p(\tau^{m+1})$ and $B_c(\tau^{m+1})$;
        \item[2.2.] Compute $u_0^{m+1}$ and $v_0^{m+1}$ by performing Step 3--9 of Algorithm 1;
        \item[2.3.] Compute $u_{n+1}^{m+1} = \kappa S_{\text{int}} e^{x_{max}}$ and $v_{n+1}^{m+1} = 0$;
        \item[2.4.] Apply the conversion-callability-puttability constraints on $u_{n+1}^{m+1}$ and $v_{n+1}^{m+1}$.
        \item[2.5.] if $\tau^{m+1} \in \mathcal{T}_{coupon}$
        \begin{enumerate}
            \item[2.5.1.] $u_{n+1}^{m+1} \leftarrow u_{n+1}^{m+1} + K$;
            \item[2.5.2.] $v_{n+1}^{m+1} \leftarrow v_{n+1}^{m+1} + K$;
        \end{enumerate}
        \item[2.6.] Compute $\bsu^{m+1}$ and $\bsv^{m+1}$ by performing Step 3--9 of Algorithm 2;
    \end{enumerate}
\end{enumerate}

For increased stability, it is possible to initiate the time integration using Rannacher's step~\cite{Rannacher1984}. In this case, for $m = 0$, the solutions $\bsu^1$, $\bsv^1$, $u_0^1$, $u_{n+1}^1$, $v_0^1$, and $v_{n+1}^1$ are computed using the initial conditions using Step 2.1--2.6 but with a smaller time step than $\Delta \tau$ (e.g, $\Delta \tau_{rann} = \Delta \tau/n_{rann}$, where $1 < n_{rann} \in \mathbb{N}$). Since we did not see stability issues when $\theta = 1/2$, we did not implement Rannacher's step to obtain numerical results presented in Section~\ref{sec:numresults}.

\section{Numerical solution of the TF model}
\label{sec:numresults}

\subsection{Comparison with FDM}

In this section, we present numerical results obtained from the FEM and time integration method discussed in Sections~\ref{sec:FEM} and~\ref{sec:timeintegral}. The modeling and computational parameters are summarized in Table~\ref{tab:parameter}, taken from~\cite{Ayache2003,Forsyth2002}. The numerical solution is computed for $x \in [-6,2]$, so that the left boundary is sufficiently close to $0$ in the $S$-space. We compare the numerical results with a second-order finite difference method (FDM), combined with the $\theta$-scheme for time integration. To have a fair comparison, we set the number of unknowns in both methods to be equal. For instance, the same meshing can be used for P1-FEM and FDM. For P2-FEM, due to additional unknowns at the midpoint of each element, we double the number of gridpoints in the FDM meshing. Throughout, $n_E$ and $n_t = n_{\tau}$ denote the number of elements and time steps, respectively. Thus, for P1-FEM, the number of unknowns (nodal points) is $n = n_E - 1$ and, for P2-FEM, $n = 2n_E - 1$. For the $\theta$-scheme, we set $\theta = \frac{1}{2}$ (Crank-Nicolson).

\begin{table}[!h]
\caption{Modeling and computational parameters} \label{tab:parameter}
\begin{center}
\begin{tabular}{ |p{6cm}||p{5.5cm}|  }
 \hline
 \textbf{Parameter} & \textbf{Value}\\
 \hline
Time to maturity $T$ & 5 years \\
Conversion & 0 to 5 years into $k$ shares \\
Conversion ratio \textit{$k$}   & 1.0 \\
Face Value $F$ & \$100 \\
Clean call price $B_c$ & \$110, from Year 3 to Year 5 ($t \in (3,5]$) \\
Clean put price $B_p$ & \$105, during Year 3 ($t \in (2,3]$)\\
Coupon payments $K$ & \$4.0\\
Coupon dates & .5, 1.0, 1.5, ... ,5.0 (semiannual)\\
Risk-free interest rate \textit{r} & 5\% or 0.05\\
Credit risk $r_c$ & 2\% or 0.02\\
Volatility $\sigma$ & 20\% or 0.2\\
Underlying stock price at $t = 0$, $S_{int}$  & \$100\\
Penalty parameter $\rho$ & $10^{12}$\\
Newton-Raphson's method tolerance  $tol$ & $10^{-12}$\\
\hline
\end{tabular}
\end{center}
\end{table}

\begin{figure}[H]
\centering
\includegraphics[width=0.49\textwidth]{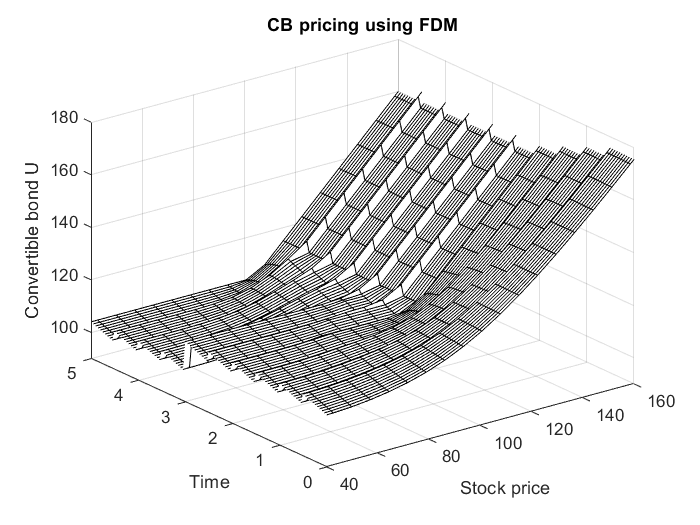}
\includegraphics[width=0.49\textwidth]{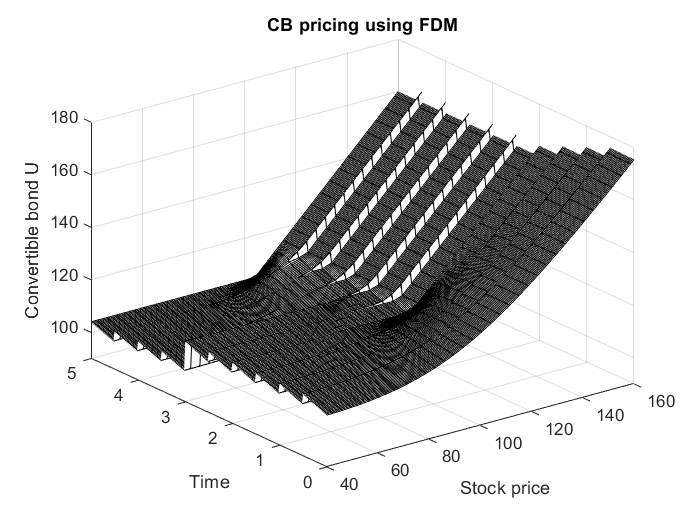}
\caption{Finite difference solution of the penalty TF model over time $t \in [0,5]$, with $r = 0.05$, $r_c = 0.02$, $\sigma = 0.2$, $F = \$100$, $K = \$4$, $\rho = 10^{12}$;  Left figure: $n = 100$, $n_t = 100$; Right figure: $n = 200$, $n_t = 200$.} 
\label{fig:FDM_surfaces}
\end{figure}

\begin{figure}[H]
\centering
\includegraphics[width=0.49\textwidth]{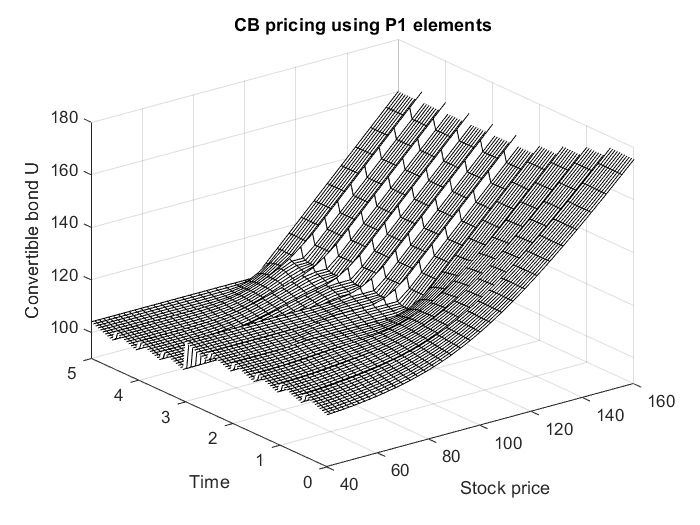}
\includegraphics[width=0.49\textwidth]{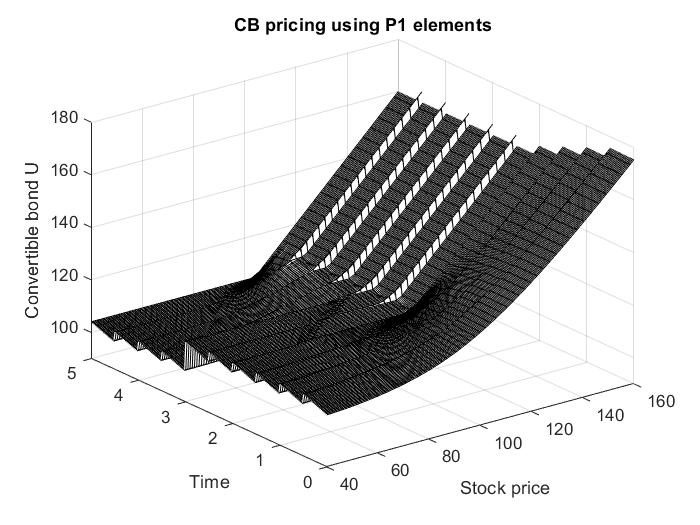}
\caption{P1-FEM solution of the penalty TF model over time $t \in [0,5]$, with $r = 0.05$, $r_c = 0.02$, $\sigma = 0.2$, $F = \$100$, $K = \$4$, $\rho = 10^{12}$;  Left figure: $n_E = 100$, $n_t = 100$; Right figure: $n_E = 200$, $n_t = 200$.}
\label{fig:P1_surfaces}
\end{figure}

\begin{figure}[H]
\centering
\includegraphics[width=0.49\textwidth]{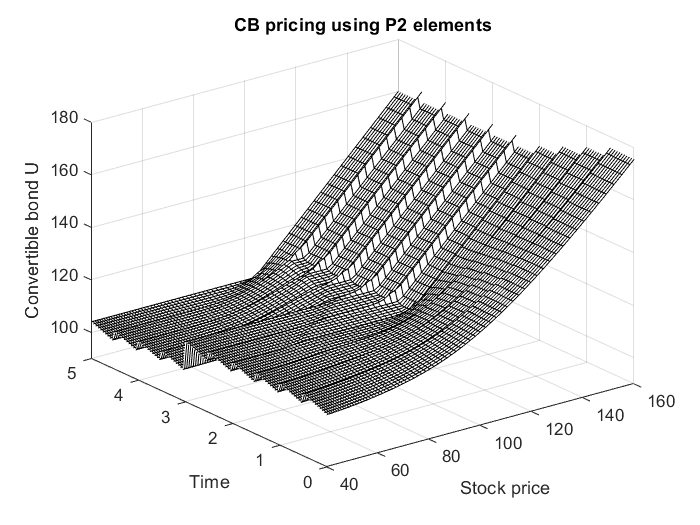}
\includegraphics[width=0.49\textwidth]{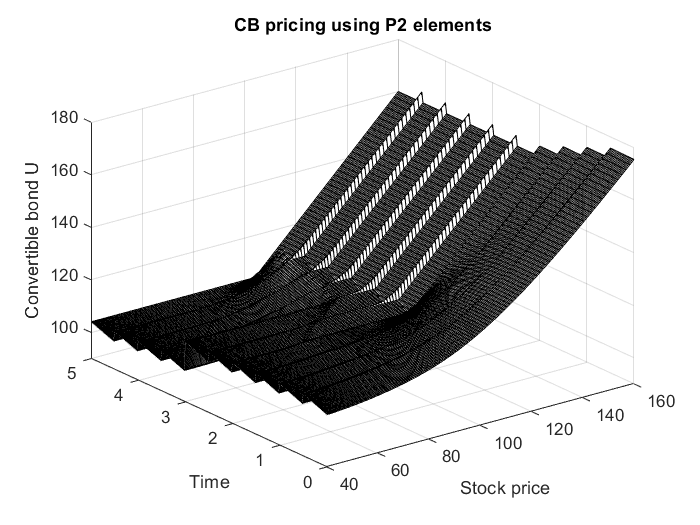}
\caption{P2-FEM solution of the penalty TF model at $t \in [0,5]$, with $r = 0.05$, $r_c = 0.02$, $\sigma = 0.2$, $F = \$100$, $K = \$4$, $\rho = 10^{12}$;  Left figure: $n_E = 100$, $n_t = 100$; Right figure: $n_E = 200$, $n_t = 200$.}
\label{fig:P2_surfaces}
\end{figure}

\begin{figure}[H]
\centering
\includegraphics[width=0.49\textwidth]{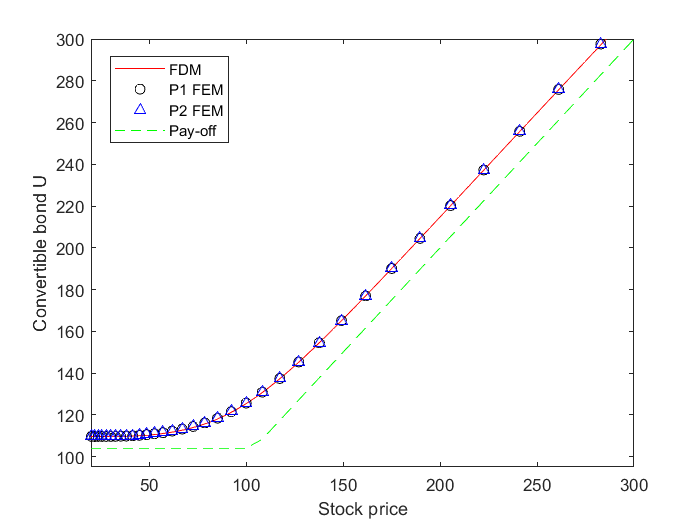}
\includegraphics[width=0.49\textwidth]{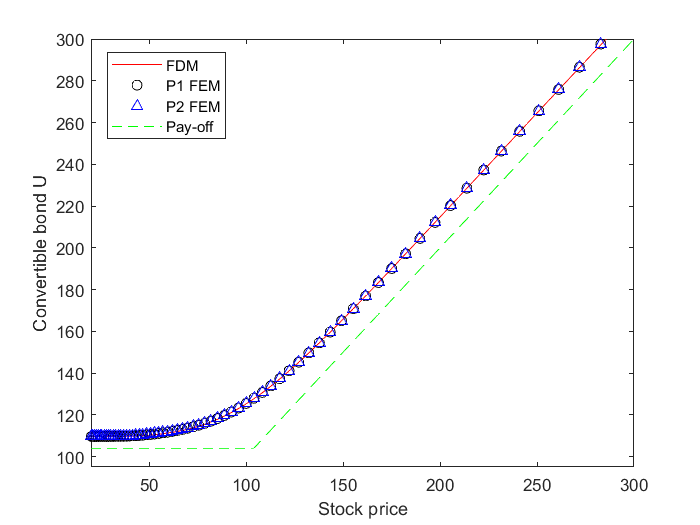}
\caption{Solution of the TF model at $t = 0$, with $r = 0.05$,$r_c = 0.02$, $\sigma = 0.2$, $F = \$100$, $K = \$4$, $\rho = 10^{12}$;  Left figure: for P1-FEM and P2-FEM, $n_E = 100$, $n_t = 100$, and for FDM $n = 200$; Right figure: for P1-FEM and P2-FEM, $n_E = 200$, $n_t = 200$, for FDM, $n = 400$.}
\label{fig:first_comparison}
\end{figure}

\begin{figure}[H]
\centering
\includegraphics[width=0.49\textwidth]{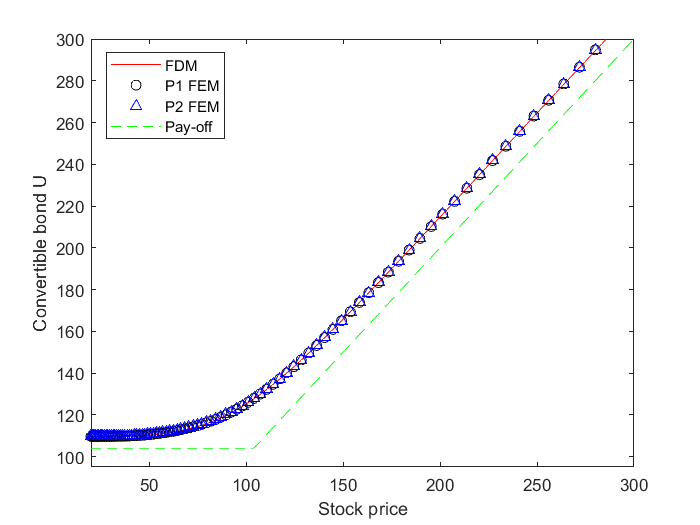}
\includegraphics[width=0.49\textwidth]{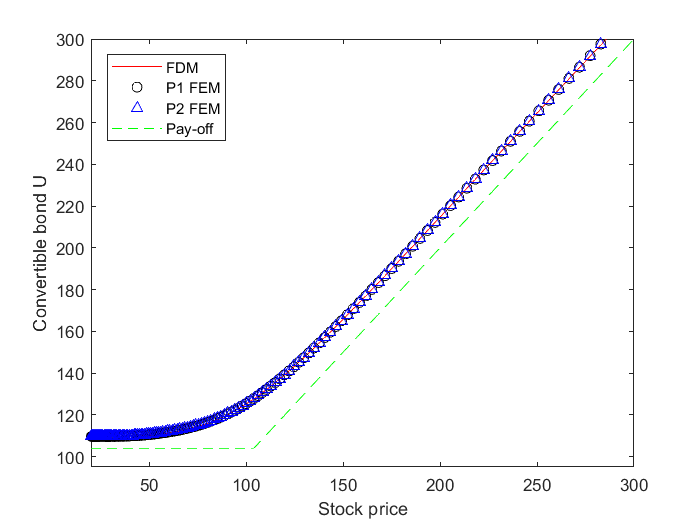}
\caption{Solution of the TF model at $t = 0$, with $r = 0.05$,$r_c = 0.02$, $\sigma = 0.2$, $F = \$100$, $K = \$4$, $\rho = 10^{12}$;  Left figure: for P1-FEM and P2-FEM, $n_E = 400$, $n_t = 400$, for FDM $n = 800$; Right figure: for P1-FEM and P2-FEM, $n_E = 800$, $n_t = 800$, for FDM $n = 1600$.}
\label{fig:second_comparison}
\end{figure}

Figures~\ref{fig:FDM_surfaces},~\ref{fig:P1_surfaces}, and~\ref{fig:P2_surfaces} show surfaces of the CB price $U$ for $40 \le S \le 160$, computed using $n_E = $ 100 and 200 and with $n_t = 200$. We note here that the steps in the solution surfaces correspond to the coupon payment time. Qualitatively, the surfaces indicate no noticeable difference between solutions of FDM and P1- or P2-FEM. That the FDM and FEM are qualitatively not distinguishable can be seen from Figure~~\ref{fig:first_comparison} and~\ref{fig:second_comparison}, where the solutions at $t = 0$ are plotted.

For convergence(mesh-independence) and comparison of the FE solutions with the FD solutions obtained using various spatial meshes, we present the bond price $U$ at the initial time $t = 0$ and $S = 100$ in Table~\ref{tab:Uprice1} computed using FDM, P1-FEM, and P2-FEM, with $n_E = n_t$ to keep the ratio $\Delta \tau/h$ constant ($\Delta \tau/h = 0.625 < 1$).
The standard stability condition for the convection-diffusion equation discretized by FDM with explicit scheme ($\theta = 0$) requires that
$$
  \Delta \tau \le C_{\theta = 0}  \min\left( \frac{h}{|a|}, \frac{h^2}{2\epsilon} \right), \quad 0 < C_{\theta = 0} \le 1,
$$
where, for the TF model, $a = r - \sigma^2/2$ and $\epsilon = \sigma^2/2$. With $n_E = n_t$, there exists however some constant $n_E^*$ such that the above stability condition is violated for $n_E > n_E^*$. Instability however was not observed in our numerical simulation with $\theta = 1/2$. Results in Table~\ref{tab:Uprice1} suggest that both P2-FEM and FDM solutions converge to a value around 124.78 (up to 2 decimal places) as the time-spatial mesh is refined. The P1-FEM solution exhibits a similar convergence towards 124.78 but at lower rate. As, can be observed that our FE solutions compared favorably with the FD solutions across these spatial meshes.
\begin{table}[h!]
\caption{Convertible bond price at $t=0$ and $S = 100$ computed by P1-FEM, P2-FEM, and FDM, with $n_E = n_t$} \label{tab:Uprice1}
\begin{center}
\begin{tabular}{cccc} \hline
$n_E$ &  P1-FEM  & P2-FEM  &  FDM \\ \hline \hline
100 & 124.422 &  124.846 & 124.991 \\
200 & 124.653 &  124.820 & 124.848 \\
400 & 124.740 &  124.814 & 124.814 \\
600 & 124.754 &  124.781 & 124.792 \\
800 & 124.742 &  124.781 & 124.790 \\
1000 & 124.746 & 124.775 & 124.780 \\ 
1200 & 124.758 & 124.777 & 124.781 \\ \hline
\end{tabular}
\end{center}
\end{table}

\subsection{The Greeks}

In the literature, because FDM is mostly used to compute the solutions, the calculation of the Greeks is done approximately using finite differencing~\cite{Christara2022,Forsyth2002} at grid points. Finite difference approaches can also in principle be used to approximate the Greeks, once the solutions are available via FEM. In this paper, we shall compute some of the Greeks, i.e., the Delta ($\Delta$) and the Gamma ($\Gamma$), directly from the finite-element approximation functions. We note, however, that since the approximating P1 FE function is piece-wise continuous and non-differentiable at the element boundaries,
we can in this case only compute the Greeks at any point inside the elements.

For P1-FEM, the approximating function is piece-wise linear continuous. Therefore, $\Delta$ can be explicitly computed by differentiating the finite element solution. In the element $\Omega_j$, the FEM solution is given by
\begin{align}
U_h(x) = - u_{j-1} (x-x_j)/h + u_j (x-x_{j-1})/h, \quad x \in \Omega_j.  \label{eq:Delta}
\end{align}
Therefore, at an instant time $t$,
$$
  \Delta(S_{j-1/2}) = \frac{\partial U_h}{\partial S}\bigg|_{S_{j-1/2}} = \frac{S_{\text{int}}}{S_{j-1/2}} \frac{\partial U_h}{\partial x} \bigg|_{x_{j-1/2}} = \frac{S_{\text{int}}}{hS_{j-1/2}}(u_j - u_{j-1})
$$
after using the change of variables, $S_{j-1/2} = S_{\text{int}} e^{x_{j-1/2}}$, $x_{j-1/2} = (x_{j-1} + x_j)/2$. 

The Greeks corresponding to higher-order derivatives however vanish. In this case, we have to resort to finite differencing, using the FEM solutions at nodal points. For instance,
$$
\frac{\partial^2 U_h}{\partial x^2} \bigg|_{x_{j-1}} \simeq (u_j - 2u_{j-1} + u_{j-2})/h^2.
$$
Hence,
\begin{align}
   \Gamma(S_{j-1}) = \frac{\partial^2 U_h}{\partial S^2} \bigg|_{S_{j-1}} = \frac{S_{\text{int}}^2}{S^2} \frac{\partial^2 U_h}{\partial x^2}\bigg|_{x_{j-1}} = \frac{S_{\text{int}}^2}{h^2S^2} (u_j - 2u_{j-1} + u_{j-2}).
\end{align}

For P2-FEM, the approximating FE function is differentiable, which in the element $\Omega_j$ reads
\begin{align}
U_h(x) = 2 u_{j-1} (x-x_{j-\frac{1}{2}})(x-x_j)/h^2 - 4 u_{j-1/2} (x-x_{j-1})(x-x_j)/h^2 + 2 u_j (x-x_{j-1})(x-x_{j-\frac{1}{2}})/h^2, \notag
\end{align}
which, after differentiation at $x_{j-1/2}$, yields
\begin{align}
  \frac{\partial U_h}{\partial x} \bigg|_{x_{j-1/2}} = (u_j - u_{j-1})/h \text{ and }
  \frac{\partial^2 U_h}{\partial x^2} \bigg|_{x_{j-1/2}} &= 4(u_j - 2u_{j-1/2} + u_{j-1})/h^2. \notag
\end{align}
This in turn results in $\Delta(S_{j-1/2})$ in~\eqref{eq:Delta} and
\begin{align}
   \Gamma(S_{j-1/2}) =  \frac{4S_{\text{int}}^2}{h^2S^2} (u_j - 2u_{j-1/2} + u_{j-1}).
\end{align}

\begin{figure}[H]
\centering
\includegraphics[width=0.49\textwidth]{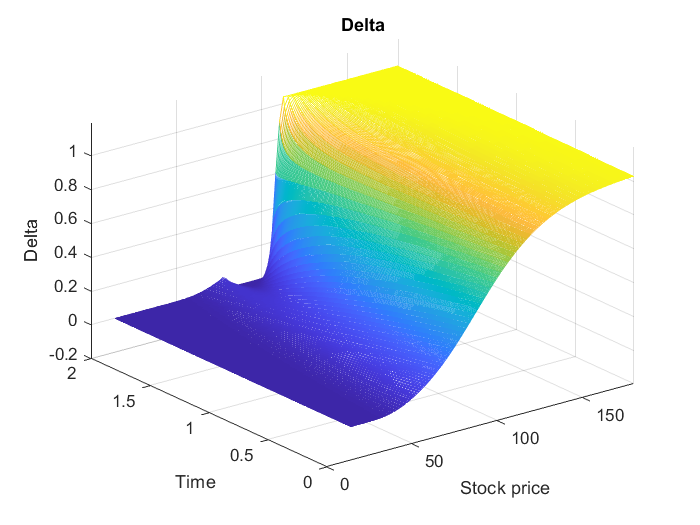}
\includegraphics[width=0.49\textwidth]{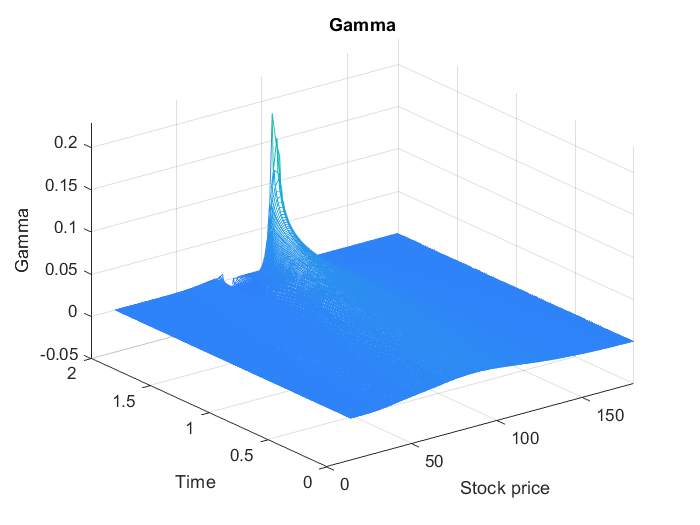}
\caption{Greeks by with the parameters $t = [0,2]$, $r = 0.05$, $r_c = 0.02$, $\sigma = 0.2$, $F = \$100$, $K = \$4$, $\rho = 10^{12}$, $n_E = 100$, $n_t = 600$ by P2-FEM;  Left figure: Delta $\Delta$; Right figure: Gamma $\Gamma$.}
\label{fig:greeks}
\end{figure}

An example of Greeks computation results is shown in Figure~\ref{fig:greeks} for $\Delta$ and $\Gamma$ during the first  two years ($t \in [0,2]$). The corresponding Greeks are computed using P2-FEM for $\Delta$ and $\Gamma$ at $S_{j-1/2}$.

Visible peaks and spikes in the solution surfaces are due to the constraints and coupon payment, at which point the solution is not differentiable; Delta and Gamma, however,  are smooth at the locations where the inequality constraints are not imposed. 

\subsection{Accuracy of the approximations}

As a way to measure the accuracy of our FEM approach, we consider a linear model problem~\cite{mms} without constraints (and hence no penalty term) and coupon payment corresponding to the exact solution
\begin{align}
U(x,\tau) &= S_{int}^2e^{2x}\sqrt{S_{int}e^x} - Fe^{-r\tau}\sqrt{S_{int}e^x}, \notag\\
V(x,\tau) &= S_{int}^2e^{2x}\sqrt{S_{int}e^x} - Fe^{-r\tau}\sqrt{S_{int}e^x} + x^2\tau, \notag
\end{align}
where $\tau \in (0,1)$. This manufactured solution corresponds to the initial conditions
\begin{eqnarray}
\begin{cases}
\displaystyle U(x,0) =S_{int}^2e^{2x}\sqrt{S_{int}e^x} - F\sqrt{S_{int}e^x}, \\
\displaystyle V(x,0) = S_{int}^2e^{2x}\sqrt{S_{int}e^x} - F\sqrt{S_{int}e^x},
\end{cases}  \notag
\end{eqnarray}
the boundary conditions
\begin{eqnarray}
\begin{cases}
\displaystyle U(0,\tau) = S_{int}^2\sqrt{S_{int}} - Fe^{-r\tau}\sqrt{S_{int}},\\
\displaystyle V(0,\tau)= S_{int}^2\sqrt{S_{int}} - Fe^{-r\tau}\sqrt{S_{int}},
\end{cases}\begin{cases}
\displaystyle U(1,\tau) = S_{int}^2e^2\sqrt{S_{int}e} - Fe^{-r\tau}\sqrt{S_{int}e}, \\
\displaystyle V(1,\tau) = S_{int}^2e^2\sqrt{S_{int}e} - Fe^{-r\tau}\sqrt{S_{int}e} +\tau,
\end{cases} \notag
\end{eqnarray}
and additional forcing terms
\begin{align}
\displaystyle f_1 &= U_{\tau} - \frac{\sigma^2}{2}U_{xx}- (r- \frac{\sigma^2}{2})U_x +rU +r_cV, \notag \\
\displaystyle f_2 &= V_{\tau} - \frac{\sigma^2}{2}V_{xx}- (r- \frac{\sigma^2}{2})V_x +(r +r_c)V, \notag
\end{align} 
in the CB and COCB PDE, respectively.  Applying FEM and the $\theta$-scheme leads to the numerical procedure
\begin{align}
     A_{11}\bsu^{m+1} &= \widetilde{A}_{11}\bsu^m - A_{12} \bsv^{m+1} + \widetilde{A}_{12} \bsv^m + \theta \Delta \tau \boldsymbol{\beta}_1^{m+1} +  (1-\theta) \Delta \tau \boldsymbol{\beta}_1^{m} + \hat{\bsb}^m_{M,u} - \hat{\bsb}^{m+1}_{M,u} +   \theta \Delta\tau f_1 + (1-\theta)\Delta\tau f_1, \notag\\
     A_{22}\bsv^{m+1} &= \widetilde{A}_{22}\bsv^m + \theta \Delta \tau \boldsymbol{\beta}_2^{m+1} +  (1-\theta) \Delta \tau \boldsymbol{\beta}_2^{m} + \hat{\bsb}^m_{M,v} - \hat{\bsb}^{m+1}_{M,v} +\theta\Delta\tau f_2 + (1-\theta) \Delta\tau f_2.  \notag
\end{align}

\begin{figure}[H]
\centering
\includegraphics[width=0.49\textwidth]{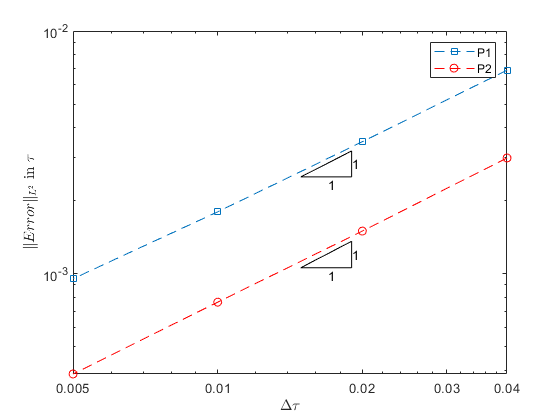}
\includegraphics[width=0.49\textwidth]{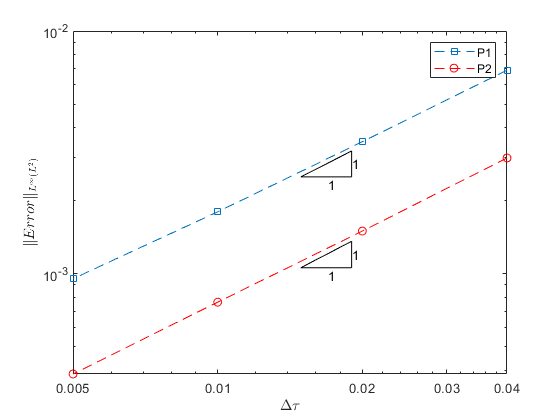}
\caption{Error estimates of the model problem in $\tau \in [0,1]$, with $r = 0.05$,$r_c = 0.02$, $\sigma = 0.2$,  $h_{P1} = 3\times 10^{-4}$, and $h_{P2} = 10^{-3}$.} 
\label{fig:error1}
\end{figure}

\begin{figure}[H]
\centering
\includegraphics[width=0.49\textwidth]{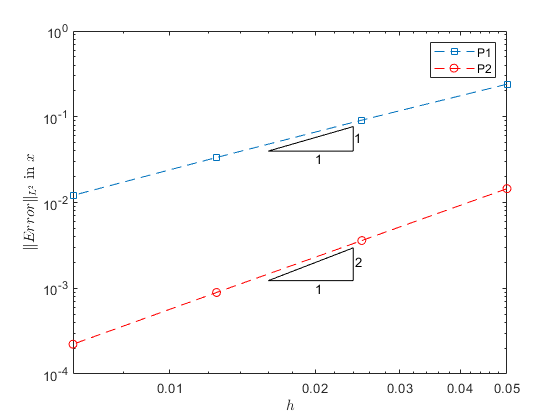}
\includegraphics[width=0.49\textwidth]{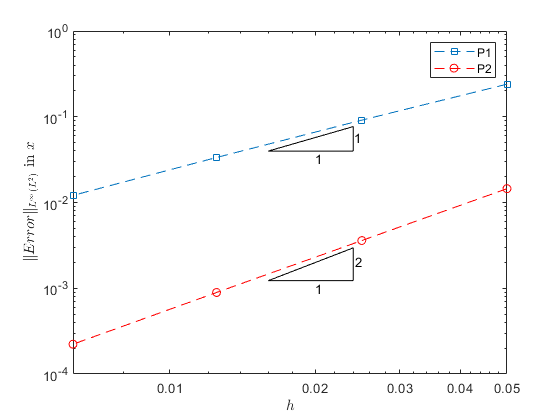}
\caption{Error estimates of the MMS model in $x \in [0,1]$, with $r = 0.05$,$r_c = 0.02$, $\sigma = 0.2$, and $\Delta\tau = 10^{-4}$.} 
\label{fig:error2}
\end{figure}

Calculated errors are presented Figure~\ref{fig:error1} and~\ref{fig:error2} using two measures \cite{amanbek2019priori,amanbek2020error}:  
\begin{align}
  \| Error\|_{L^2} &= \|  U(x,1)  - \bsu^{n_t}\|_{L^2},  \notag\\
   \| Error\|_{L^\infty (L^2)} &= \max_{1\le m \le n_t}\left(\| U(x,m \Delta \tau) - \bsu^m \|_{L^2} \right)\notag,
\end{align}
where $\bsu^m$ is the solution of the model problem at $\tau = m \Delta \tau$, computed by P1-FEM or P2-FEM. In Figure~\ref{fig:error1}, the errors are calculated for varying $\Delta  \tau$ and a fixed value of $h$. The errors decreases as $\Delta \tau$ is reduced to 0, with a rate that is proportional to $\Delta \tau$ (first-order convergence). This first order convergence is consistent with the result in \cite{Ayache2003}.

In Figure~~\ref{fig:error2}, the errors are calculated for varying $h$ and fixed $\Delta \tau$. The plots suggest convergence of P1-FEM and P2-FEM at the rate proportional to $h$ and $h^2$, respectively (Theoretically, e.g. for P1-FEM the convergence rate is given by the relation $\| Error\|_{L^\infty (L^2)} \le C(h+\Delta \tau)$). This convergence rate is as expected for the linear model used but cannot, however, be expected when nonlinear (e.g., penalty) terms are added. 
 
\section{Conclusion}
\label{sec:conclusion}
We presented numerical methods based on  P1 and P2 finite element methods for the TF model in convertible bond pricing, reformulated as a penalty PDE. The resultant differential algebraic equations were solved using $\theta$-scheme, with nonlinearity due to the penalty handled by Newton's method.

We reported results from the numerical simulations using the two finite element approximations and compared them with finite difference methods. Comparing results from the two methods, we observed extremely comparable results. Furthermore, we verified the advantage of using P2-FEM over P1-FEM, when it comes to accuracy of the solutions using fewer elements. 

The finite-element method described in this paper can be implemented on different models for convertible bond pricing, such as the AFV model~\cite{Ayache2003}.  Results on this will be reported in the future.

\bibliography{main}

\newcommand{\noopsort}[1]{} \newcommand{\printfirst}[2]{#1}
  \newcommand{\singleletter}[1]{#1} \newcommand{\switchargs}[2]{#2#1}
\begin{thebibliography}{10}

\bibitem{AlSaedi2018}
Y.~H. AlSaedi and G.~A. Tularam.
\newblock A review of the recent advances made in the {Black-Scholes} models
  and respective solutions methods.
\newblock {\em Journal of Mathematics and Statistics}, 14(1):29--39, January
  2018.

\bibitem{amanbek2020error}
Y.~Amanbek, G.~Singh, G.~Pencheva, and M.~F. Wheeler.
\newblock Error indicators for incompressible {Darcy} flow problems using
  enhanced velocity mixed finite element method.
\newblock {\em Computer Methods in Applied Mechanics and Engineering},
  363:112884, 2020.

\bibitem{amanbek2019priori}
Y.~Amanbek and M.~F. Wheeler.
\newblock A priori error analysis for transient problems using enhanced
  velocity approach in the discrete-time setting.
\newblock {\em Journal of Computational and Applied Mathematics}, 361:459--471,
  2019.

\bibitem{Ammann2008}
M.~Ammann, A.~Kind, and C.~Wilde.
\newblock Simulation-based pricing of convertible bonds.
\newblock {\em Journal of Empirical Finance}, 15(2):310--331, 2008.

\bibitem{Ankudinova2008}
J.~Ankudinova and M.~Ehrhardt.
\newblock On the numerical solution of nonlinear {Black-Scholes} equations.
\newblock {\em Computers and Mathematics with Applications}, 56(3):799--812,
  August 2008.

\bibitem{Ayache2003}
E.~Ayache, P.~A. Forsyth, and K.~R. Vetzal.
\newblock Valuation of convertible bonds with credit risk.
\newblock {\em The Journal of Derivatives}, 11(1):9--29, 2003.

\bibitem{BaroneAdesi2003}
G.~Barone-Adesi, A.~Bermudez, and J.~Hatgioannides.
\newblock Two-factor convertible bonds valuation using the method of
  characteristics/finite elements.
\newblock {\em Journal of Economic Dynamics and Control}, 27(10):1801--1831,
  August 2003.

\bibitem{Black1973ThePO}
F.~Black and M.~S. Scholes.
\newblock The pricing of options and corporate liabilities.
\newblock {\em Journal of Political Economy}, 81:637--654, 1973.

\bibitem{Brennan}
M.~J. Brennan and E.~S. Schwartz.
\newblock Convertible bonds: Valuation and optimal strategies for call and
  conversion.
\newblock {\em The Journal of Finance}, 32:1699--1715, 1977.

\bibitem{Brennan2}
M.~J. Brennan and E.~S. Schwartz.
\newblock Analyzing convertible bonds.
\newblock {\em Journal of Financial and Quantitative Analysis}, 15:907--929,
  1980.

\bibitem{ern2016}
D.~{\v{C}}ern{\'{a}}.
\newblock Numerical solution of the {Black-Scholes} equation using cubic spline
  wavelets.
\newblock In G.~Venkov et~al., editor, {\em {AIP} Conference Proceedings 1789},
  page 030001. {AIP} Publishing, 2016.

\bibitem{Christara2022}
C.~C. Christara and R.~Wu.
\newblock Penalty and penalty-like methods for nonlinear {HJB} {PDEs}.
\newblock {\em Applied Mathematics and Computation}, 425:127015, 2022.

\bibitem{deFrutos2005}
J.~de~Frutos.
\newblock A finite element method for two factor convertible bonds.
\newblock In {\em Numerical Methods in Finance}, pages 109--128. Springer {US},
  2005.

\bibitem{Dremkova2011}
E.~Dremkova and M.~Ehrhardt.
\newblock A high-order compact method for nonlinear {Black-Scholes} option
  pricing equations of {American} options.
\newblock {\em International Journal of Computer Mathematics},
  88(13):2782--2797, 2011.

\bibitem{FLETCHER1983225}
C.~A.~J. Fletcher.
\newblock The group finite element formulation.
\newblock {\em Computer Methods in Applied Mechanics and Engineering},
  37(2):225--244, 1983.

\bibitem{Forsyth2002}
P.~A. Forsyth and K.~R. Vetzal.
\newblock Quadratic convergence for valuing {American} options using a penalty
  method.
\newblock {\em {SIAM} Journal on Scientific Computing}, 23(6):2095--2122,
  January 2002.

\bibitem{Forsyth1999}
P.~A. Forsyth, K.~R. Vetzal, and R.~Zvan.
\newblock A finite element approach to the pricing of discrete lookbacks with
  stochastic volatility.
\newblock {\em Applied Mathematical Finance}, 6(2):87--106, 1999.

\bibitem{hull2003options}
J.~C. Hull.
\newblock {\em Options, Futures, and Other Derivatives}.
\newblock Pearson Education India, 2003.

\bibitem{Jonatha-1977}
J.~E.~Ingersoll Jr.
\newblock A contingent-claims valuation of convertible securities.
\newblock {\em Journal of Financial Economics}, 4:289--321, May 1977.

\bibitem{Kovalov2008}
P.~Kovalov and V.~Linetsky.
\newblock Valuing convertible bonds with stock price, volatility, interest
  rate, and default risk.
\newblock {\em {SSRN} Electronic Journal}, 2008.

\bibitem{Kythe-2004}
P.~K. Kythe and D.~Wei.
\newblock {\em An Introduction to Linear and Nonlinear Finite Element Analysis:
  A Computational Approach}, volume~57.
\newblock The American Society of Mechanical Engineers, 2004.

\bibitem{Lin2020}
S.~Lin and S.-P. Zhu.
\newblock Numerically pricing convertible bonds under stochastic volatility or
  stochastic interest rate with an {ADI}-based predictor-corrector scheme.
\newblock {\em Computers and Mathematics with Applications}, 79(5):1393--1419,
  2020.

\bibitem{Milanov2012}
K.~Milanov and O.~Kounchev.
\newblock Binomial tree model for convertible bond pricing within equity to
  credit risk framework.
\newblock {\em {SSRN} Electronic Journal}, 2012.

\bibitem{Rannacher1984}
R.~Rannacher.
\newblock Finite element solution of diffusion problems with irregular data.
\newblock {\em Numerische Mathematik}, 43(2):309--327, 1984.

\bibitem{mms}
P.~J. Roache.
\newblock Code verification by the method of manufactured solutions.
\newblock {\em Journal of Fluids Engineering}, 124:4--10, 2001.

\bibitem{handbook}
J.~De Spiegeleer, W.~Schoutens, and C.~Van Hulle.
\newblock {\em The Handbook of Hybrid Securities: Convertible Bonds, CoCo Bonds
  and Bail-In}.
\newblock The Wiley Finance Series. Wiley, 1st edition, 2014.

\bibitem{wiley}
J.~De Spiegeleer, W.~Schoutens, and P.~Jabre.
\newblock {\em The Handbook of Convertible Bonds: Pricing, Strategies and Risk
  Management}.
\newblock The Wiley Finance Series. Wiley, 1st edition, 2011.

\bibitem{vcb}
K.~Tsiveriotis and C.~Fernandes.
\newblock Valuing convertible bonds with credit risk.
\newblock {\em The Journal of Fixed Income}, 8:95--102, September 1998.

\end{thebibliography}
\bibliographystyle{plain}

\end{document}